\documentclass[12pt]{article}

\input epsf

\renewcommand{\arraystretch}{2.0} 

\def\dfrac#1#2{{\displaystyle {#1 \over #2}}}
\def\dsum{\mathop{\displaystyle \sum }}

\def\simge{\mathrel{\rlap{\raise 0.511ex \hbox{$>$}}{\lower 0.511ex \hbox{$\sim$}}}}
\def\simle{\mathrel{\rlap{\raise 0.511ex \hbox{$<$}}{\lower 0.511ex \hbox{$\sim$}}}} 
\def\slash#1{\setbox0=\hbox{$#1$}\dimen0=\wd0                    
      \setbox1=\hbox{/} \dimen1=\wd1 \ifdim\dimen0>\dimen1
      \rlap{\hbox to \dimen0{\hfil/\hfil}} #1                        \else                                       
      \rlap{\hbox to \dimen1{\hfil$#1$\hfil}}
      /   \fi}                                         

\newcommand{\be}{\begin{equation}}
\newcommand{\ee}{\end{equation}}
\newcommand{\bea}{\begin{eqnarray}}
\newcommand{\eea}{\end{eqnarray}}
\newcommand{\nn}{\nonumber}
\newcommand{\as}{\alpha_{ s}}

\newcommand{\Heff}{{\cal H}_{ eff}}
\newcommand{\DB}{\Delta B}

\newcommand{\msb}{\mbox{NDR-}\overline{\rm{MS}}}

\newcommand{\ep}{\varepsilon}

\newcommand{\rp}{\tau(B^+)/\tau(B_d)}
\newcommand{\rs}{\tau(B_s)/\tau(B_d)}
\newcommand{\rl}{\tau(\Lambda_b)/\tau(B_d)}

\begin{document}

\begin{titlepage}
\begin{flushright}
RM3-TH/02-3\\
ROME1-1333/02\\
SHEP 02/04
\end{flushright}
\vskip 2.4cm
\begin{center}
{\Large \bf Lifetime Ratios of Beauty Hadrons at the Next-to-Leading Order in
QCD \\  \vspace{0.2cm}}

\vskip1.3cm 
{\large\bf E.~Franco$^a$, V.~Lubicz$^b$, F.~Mescia$^c$ and C.~Tarantino$^b$}\\

\vspace{1.cm}
{\normalsize {\sl 
$^a$ Dip. di Fisica, Univ. di Roma ``La Sapienza" and INFN,
Sezione di Roma,\\ P.le A. Moro 2, I-00185 Rome, Italy \\
\vspace{.25cm}
$^b$ Dip. di Fisica, Univ. di Roma Tre and INFN,
Sezione di Roma III, \\
Via della Vasca Navale 84, I-00146 Rome, Italy \\
\vspace{.25cm}
$^c$ Dept. of Physics and Astronomy, Univ. of Southampton, \\
Highfield, Southampton, SO17 1BJ, U.K.
}}
\vspace{.25cm}
\vskip1.5cm
\abstract{
We compute the next-to-leading order QCD corrections to spectator effects in 
the lifetime ratios of beauty hadrons. With respect to previous calculations, 
we take into account the non vanishing value of the charm quark mass. We obtain
the predictions $\tau(B^+)/\tau(B_d) = 1.06 \pm 0.02$, $\tau(B_s)/\tau(B_d)= 
1.00 \pm 0.01$ and $\tau(\Lambda_b)/\tau(B_d) = 0.90 \pm 0.05$, in good
agreement with the experimental results. In the case of $\tau(B_s)/\tau(B_d)$
and $\tau(\Lambda_b)/\tau(B_d)$, however, some contributions, which either 
vanish in the vacuum insertion approximation or represent a pure NLO 
corrections, have not been determined yet.}
\end{center}
\vspace*{1.cm}
\end{titlepage}

\setcounter{footnote}{0}
\setcounter{equation}{0}

\section{Introduction}
\label{sec:intro}
The inclusive decay rates of beauty hadrons can be computed by expanding the
amplitudes in increasing powers of $\Lambda_{QCD}/m_b$~\cite{Khoze:1987fa,ope} 
and using the assumption of quark-hadron duality. The leading term in this 
expansion reproduces the predictions of the na\"\i ve quark spectator model, 
and the first correction of ${\cal O}(1/m_b)$ is absent in this case. 

Within this theoretical framework, up to terms of ${\cal O}(1/m_b^2)$ only the 
$b$ quark enters the short-distance weak decay, while the light quarks in the 
hadron interact through soft gluons only. For this reason, the lifetime ratios
of beauty hadrons are predicted to be unity at this order, with corrections
which are at most of few percent.

Spectator contributions, which are expected to be mainly responsible for the 
lifetime differences of beauty hadrons, only appear at ${\cal O}(1/m_b^3)$ in 
the Heavy Quark Expansion (HQE). These effects, although suppressed by powers 
of $1/m_b$, are enhanced, with respect to leading contributions, by a 
phase-space factor of $16\pi^2$, being $2\to 2$ processes instead of $1\to 3$ 
decays. Indeed, the inclusion of these corrections at the leading order (LO) in
QCD has allowed to reproduce the observed pattern of the lifetime 
ratios~\cite{primisp,NS}. 

In this paper we compute the next-to-leading order (NLO) QCD corrections to
spectator effects in the lifetime ratios of beauty hadrons. In the limit of 
vanishing charm quark mass, these corrections have been already computed in 
ref.~\cite{noantri}, and we refer to this paper for many details of the NLO
calculation. We find that the inclusion of the NLO corrections improves the 
agreement with the experimental measurements. Moreover, it increases the
accuracy of the theoretical predictions by reducing significantly the 
dependence on the operator renormalization scale.

After our calculation was completed, the NLO corrections to spectator effects in
the ratio $\rp$ have been also presented in ref.~\cite{bucha}. We agree with 
their results. With respect to ref.~\cite{bucha}, we also perform in this paper 
the NLO calculation of the Wilson coefficients entering the HQE of the ratios 
$\rs$ and $\rl$. 

By using the lattice determinations of the relevant hadronic matrix 
elements~\cite{DiPierro98}-\cite{APE}, we have performed a theoretical estimate
of the lifetime ratios and obtained the NLO predictions
\be
\label{eq:nlores}
\frac{\tau(B^+)}{\tau(B_d)} = 1.06 \pm 0.02 \, ,  \qquad
\frac{\tau(B_s)}{\tau(B_d)} = 1.00 \pm 0.01 \, ,  \qquad
\frac{\tau(\Lambda_b)}{\tau(B_d)} = 0.90 \pm 0.05\, .
\ee
These estimates must be compared with the experimental 
measurements~\cite{ckmwork}
\be
\label{eq:rexp}
\frac{\tau(B^+)}{\tau(B_d)}=1.074 \pm 0.014 \, , \qquad
\frac{\tau(B_s)}{\tau(B_d)}=0.948 \pm 0.038 \, , \qquad
\frac{\tau(\Lambda_b)}{\tau(B_d)}=0.796 \pm 0.052\, .
\ee
The theoretical predictions turn out to be in good agreement with the 
experimental data. As we will discuss below, however, in the case of $\rs$ and 
$\rl$ the theoretical predictions can be further improved, since some 
contributions, which either vanish in the vacuum insertion approximation (VIA)
or represent a pure NLO correction, are still lacking. With respect to the 
results of ref.~\cite{noantri}, we find that the NLO charm quark mass 
corrections are rather large for some of the Wilson coefficients. Nevertheless, 
the total effect of these contributions on the lifetime ratios is numerically 
small.

An important check of the perturbative calculation performed in this paper is 
provided by the cancellation of the infrared (IR) divergences in the 
expressions of the Wilson coefficients, in spite these divergences appear in 
the individual amplitudes. The presence of these divergences explicitly shows
that the Bloch-Nordsieck theorem does not apply in non-abelian gauge 
theories~\cite{Bloch:1937pw}-\cite{Di'Lieto:1980dt}. We have also checked that 
our results are explicitly gauge invariant and have the correct ultraviolet 
(UV) renormalization-scale dependence as predicted by the known LO anomalous 
dimensions of the relevant operators.

The HQE of the lifetime ratios results in a series of local operators of
increasing dimension defined in the Heavy Quark Effective Theory (HQET). 
Indeed, renormalized operators in QCD mix with operators of lower dimension, 
with coefficients proportional to powers of the $b$-quark mass. In this case, 
the dimensional ordering of the HQE would be lost. In order to implement the
expansion, the matrix elements of the local operators should be cut-off at a
scale smaller than the $b$-quark mass, which is naturally realized in the 
HQET.\footnote{On this point, we disagree with the general statement of 
ref.~\cite{bucha} according to which the HQE can be performed in QCD also in
the presence of mixing with lower dimensional operators.}
In the case of $\rp$, only flavour non-singlet operators enter the expansion. 
For these operators, the mixing with lower dimensional operators is absent, and
the HQE can be expressed in terms of operators defined in QCD. For these
operators, the matching between QCD and HQET has been computed, at the NLO, in 
ref.~\cite{noantri}.

As mentioned before, the NLO predictions for the lifetime ratios $\rs$ and 
$\rl$ may still be improved in some ways. In particular:
\begin{itemize}
\item[-] the contributions of the four-fermion operators containing the charm 
quark field and of the penguin operator (see eqs.~(\ref{eq:opeff}) and 
(\ref{eq:openg})) have not been determined yet. These contributions, which 
either vanish in the VIA (charm operators) or represent a pure NLO correction
(penguin operator), do not affect the theoretical determination of $\rp$ and
represent only an $SU(3)$-breaking effect for $\rs$;

\item[-] the lattice determination of the hadronic matrix elements have been
performed by neglecting penguin contractions (i.e. eye diagrams), which exist
in the case of flavour singlet operators. Also in this case, the corresponding
contributions cancel in the ratio $\rp$ but affect $\rl$ and $\rs$, the latter
only through $SU(3)$-breaking effects;

\item[-] the NLO anomalous dimension of the four-fermion $DB=0$ operators in 
the HQET is still unknown. For this reason, the renormalization scale evolution 
of the $\Lambda_b$ matrix elements, computed on the lattice in the HQET at a 
scale smaller than the $b$-quark mass~\cite{DiPierro:1999tb,DiPierroproc}, has 
been only performed at the LO. For $B$ mesons, a complete NLO evolution has 
been performed by using operators defined in QCD~\cite{APE}.
\end{itemize}
For these reasons, at present, the best theoretical accuracy is achieved in the 
determination of the lifetime ratio $\rp$.

We conclude this section by presenting the plan of this paper. In 
sect.~\ref{sec:formulae} we review the basic formalism of the HQE applied to 
the lifetime ratios of beauty hadrons. Details of the NLO calculation and the 
numerical results obtained for the Wilson coefficients are given in 
sect.~\ref{sec:result}. In sect.~\ref{sec:phenom} we present the predictions 
for the lifetime ratios of beauty hadrons. Finally, we collect in the appendix
the analytical expressions for the Wilson coefficient functions.

\section{HQE for the lifetime ratios of beauty hadrons}
\label{sec:formulae}
Using the optical theorem, the inclusive decay width $\Gamma(H_b)$ of a hadron 
containing a $b$ quark can be written as
\be
\Gamma(H_b) =
\frac{1}{M_{H_b}} \mathrm{Im} \langle H_b \vert {\cal T} \vert H_b \rangle\, ,
\label{eq:master}
\ee
where the transition operator ${\cal T}$ is given by
\be 
{\cal T} = i \int d^4x \; T \left( \Heff^{\DB=1}(x) \Heff^{\DB=1}(0) \right)
\label{eq:T}
\ee
and $\Heff^{\DB=1}$ is the effective weak hamiltonian which describes $\DB=1$ 
transitions. 

By neglecting the Cabibbo suppressed contribution of $b \to u$ transitions 
($\vert V_{ub}\vert^2/\vert V_{cb}\vert^2 \sim 0.16 \lambda^2$) and terms 
proportional to $\vert V_{td}\vert/\vert V_{ts}\vert$ in the penguin sector, the
$\DB=1$ effective hamiltonian can be written in the form
\be
\label{eq:hdb1}
\Heff^{\DB=1}=
\frac{G_F}{\sqrt{2}} V^\ast_{cb} \Bigg[ 
C_1 \left(Q_1 + Q_1^c \right) + C_2 \left( Q_2 + Q_2^c \right) +
\sum_{i=3}^{6} C_i Q_i + C_{8G} Q_{8G} +
\dsum_{l=e,\mu,\tau} Q_l \Bigg] + h.c. \, .
\ee
The $C_i$ are the Wilson coefficients, known at the NLO in perturbation
theory~\cite{nlodb1a}-\cite{nlodb1c}, and the operators $Q_i$ are defined as
\be
\begin{array}{ll}
Q_1= (\bar b_i c_j)_{V-A} (\bar u_j d_i^ {\, \prime})_{V-A}\,, \qquad \qquad
& Q_2= (\bar b_i c_i)_{V-A} (\bar u_j d_j^ {\, \prime})_{V-A}\,, \\
Q^c_1= (\bar b_i c_j)_{V-A} (\bar c_j s_i^ {\, \prime})_{V-A}\,, \qquad \qquad
& Q^c_2= (\bar b_i c_i)_{V-A} (\bar c_j s_j^ {\, \prime})_{V-A}\,, \\
Q_3= (\bar b_i s_i)_{V-A} \dsum\limits_{q} (\bar q_j q_j)_{V-A}\,, \qquad \qquad
& Q_4= (\bar b_i s_j)_{V-A} \dsum\limits_{q} (\bar q_j q_i)_{V-A}\,, \\
Q_5= (\bar b_i s_i)_{V-A} \dsum\limits_{q} (\bar q_j q_j)_{V+A}\,, \qquad \qquad
& Q_6= (\bar b_i s_j)_{V-A} \dsum\limits_{q} (\bar q_j q_i)_{V+A}\,, \\
Q_{8G}= \dfrac{g_s}{8\pi^2} m_b \bar b_i \sigma^{\mu\nu} 
\left(1-\gamma^5\right) t^{a}_{ij} s_j G^{a}_{\mu\nu} \,, & Q_l=(\bar b_i
c_i)_{V-A}(\bar\nu_l l)_{V-A}\,,
\end{array}
\label{eq:operatori}
\ee
where $d^{\, \prime}= \cos\theta_c d + \sin\theta_c s$ and $s^{\, \prime}= 
-\sin\theta_c d + \cos\theta_c s$ ($\theta_c$ is the Cabibbo angle). Here and 
in the following we use the notation $(\bar q q)_{V\pm A}=\bar q \gamma_\mu(1
\pm\gamma_5) q$ and $(\bar q q)_{S\pm P}=\bar q (1\pm\gamma_5) q$. A sum over 
repeated colour indices is always understood.

\begin{figure} [t]
\begin{center}
\epsfxsize=12cm
\epsfbox{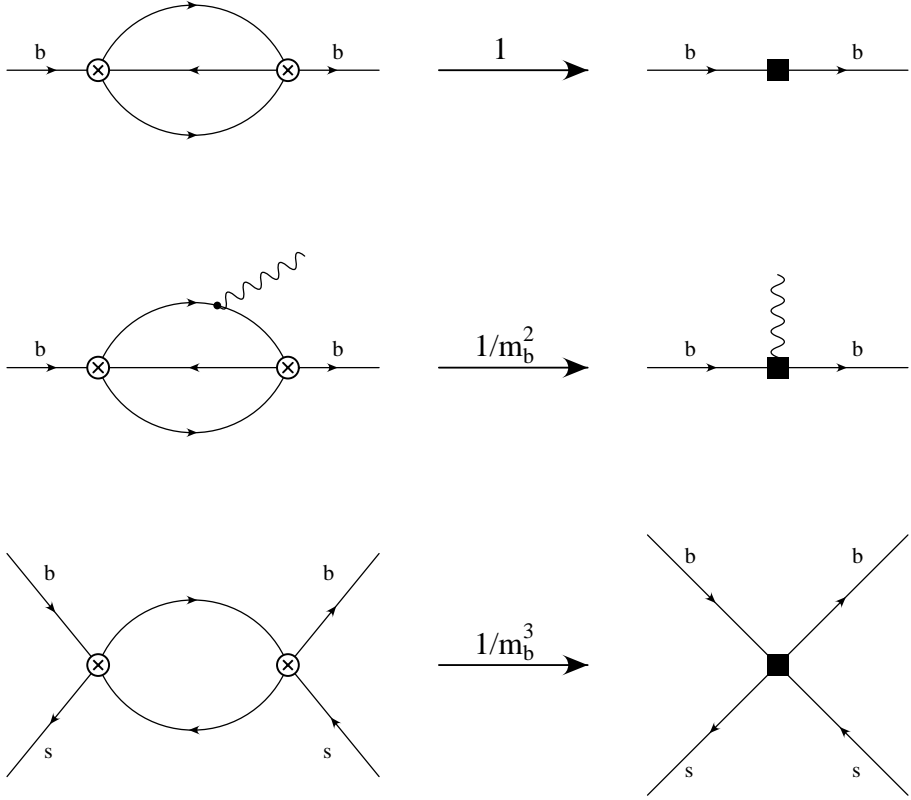}
\end{center}
\caption{\it Examples of LO contributions to the transition operator ${\cal T}$
(left) and to the corresponding local operator (right). The crossed circles 
represent the insertions of the $\DB=1$ effective hamiltonian. The black 
squares represent the insertion of a $\DB=0$ operator.}
\label{fig:tree}
\end{figure}

Because of the large mass of the $b$ quark, it is possible to construct an 
Operator Product Expansion (OPE) for the transition operator ${\cal T}$ of 
eq.~(\ref{eq:T}), which results in a sum of local operators of increasing 
dimension~\cite{Khoze:1987fa,ope}. We include in this expansion terms up to 
${\cal O}(1/m_b^2)$ plus those $1/m_b^3$ corrections that come from spectator 
effects and are enhanced by the phase space. The resulting expression for the 
inclusive width of eq.~(\ref{eq:master}) is given by   
\be
\Gamma(H_b) =
\frac{G_F^2 |V_{cb}|^2 m_b^5}{192 \pi^3}\left[
c^{(3)} \frac{\langle \bar b b \rangle_{H_b}}{2 M_{H_b}} +
c^{(5)} \frac{g_s}{m_b^2} \frac{\langle \bar b\sigma_{\mu\nu}G^{\mu\nu}b
\rangle_{H_b}}{2 M_{H_b}}+
\frac{96\pi^2}{m_b^3} \dsum_{k} c^{(6)}_k \frac{\langle O_k^{(6)}\rangle_{H_b}}
{2 M_{H_b}}\right]\, ,
\label{eq:gamma}
\ee
where $\langle\cdots\rangle_{H_b}$ denotes the forward matrix element between
two hadronic states $H_b$, defined with the covariant normalization. The
operators $O_k^{(6)}$ are a set of four-fermion, dimension-six operators to
be specified below, which represent the contribution of hard spectator effects. 
At the lowest order in QCD, the diagrams entering the calculation of 
$\Gamma(H_b)$ are shown in fig.~\ref{fig:tree}. 

The matrix elements of dimension-three and dimension-five operators, appearing 
in eq.~(\ref{eq:gamma}), can be expanded by using the HQET
\bea
\langle \bar b b\rangle_{H_b}&=&2 M_{H_b} \left( 1-\frac{\mu_\pi^2(H_b)-
\mu_G^2(H_b)}{2 m_b^2}+{\cal O}(1/m_b^3)\right)\, ,\nn \\
g_s
\langle\bar b\sigma_{\mu\nu}G^{\mu\nu} b\rangle_{H_b}&=& 2M_{H_b}
\left(2\mu^2_G(H_b)+{\cal O}(1/m_b)\right)\, .
\label{eq:hqet}
\eea
By using these expansions, we can finally compute, from eq.~(\ref{eq:gamma}),
the lifetime ratio of two beauty hadrons
\bea
\label{eq:ratio}
\dfrac{\Gamma(H_b)}{\Gamma(H_b^\prime)}&=&1-\frac{\mu_\pi^2(H_b)-
\mu_\pi^2(H_b^\prime)}{2 m_b^2}+\left(\frac{1}{2}+\frac{2 c^{(5)}}{c^{(3)}}
\right) \frac{\mu_G^2(H_b)-\mu_G^2(H_b^\prime)}{m_b^2}+\nn\\
&&\quad\dfrac{96\pi^2}{m_b^3\, c^{(3)}}\dsum\limits_{k}
c^{(6)}_k \left(\frac{\langle O^{(6)}_k\rangle_{H_b}}{2M_{H_b}}-\frac{\langle
O_k^{(6)}\rangle_{H_b^\prime}}{2M_{H_b^\prime}}\right) \, .
\eea

The dimension-six operators in eq.~(\ref{eq:ratio}), which express the hard
spectator contributions, are the four current-current operators\footnote{Note
that, with respect to ref.~\cite{noantri}, we use a different basis of 
operators.}
\be
\begin{array}{ll}
{\cal O}^q_1= (\bar b \,q)_{V-A} \, (\bar q \,b)_{V-A}\,, &
{\cal O}^q_2= (\bar b \,q)_{S-P} \, (\bar q \,b)_{S+P}\,, \\
{\cal O}^q_3= (\bar b \,t^{a} q)_{V-A} \, (\bar q \,t^{a} b)_{V-A}\,, &
{\cal O}^q_4= (\bar b \,t^{a} q)_{S-P} \, (\bar q \,t^{a} b)_{S+P}\, ,
\end{array}
\label{eq:opeff}
\ee
with $q=u,d,s,c$, and the penguin operator
\be
{\cal O}_P= (\bar b t^{a} b)_{V} \dsum\limits_{q=u,d,s,c}(\bar q t^{a} q)_{V} 
\, .
\label{eq:openg}
\ee
In these definitions, the symbols $b$ and $\bar b$ denote the heavy quark
fields in the HQET.

The Wilson coefficients $c^{(3)}$ and $c^{(5)}$, of the dimension-three and
dimension-five operators in eq.~(\ref{eq:ratio}), have been computed at the LO 
in ref.~\cite{Bigi:1992su}, while the NLO corrections to $c^{(3)}$ have been 
evaluated in \cite{gamma1}-\cite{gamma6}. The NLO corrections to $c^{(5)}$ are 
still missing. Their numerical contribution to the lifetime ratios, however, is 
expected to be negligible.

The coefficient functions of the dimension-six current-current operators 
${\cal O}^q_k$, have been computed at the LO in ref.~\cite{NS,Ura} for 
$q=u,d,s$, and in ref.~\cite{charm} for $q=c$. The coefficient function of the 
penguin operator ${\cal O}_P$ vanishes at the LO.

In this paper we have computed the NLO QCD corrections to the coefficient
functions of the operators ${\cal O}^q_k$ with $q=u,d,s$. The operators 
containing the charm quark fields contribute, as valence operators, only to the 
inclusive decay rate of $B_c$ mesons, and their contribution to non-charmed
hadron decay rates is expected to be negligible. The calculation of the NLO
corrections to these coefficient functions, as well as the NLO calculation of
the coefficient function of the penguin operator, has not been performed yet.
The non-perturbative determinations of the corresponding matrix elements are
also lacking at present. However, as we will discuss in sect.~\ref{sec:phenom},
these contributions only enter the theoretical estimates of the ratios $\rs$ and
$\rl$, and vanish in the VIA.

\section{NLO calculation of the Wilson coefficients}
\label{sec:result}
All the details of the matching procedure used to determine the Wilson 
coefficients at the NLO have been given in ref.~\cite{noantri}, where the 
calculation has been performed in the limit of vanishing charm quark mass. A 
finite value of the charm quark mass does not introduce conceptual difficulties 
in the calculation, besides requiring the evaluation of one- and two-loop 
integrals with an additional mass scale. For this reason, we only remind in 
this section the general strategy of the perturbative calculation, and refer 
the interested reader to ref.~\cite{noantri} for further details. In this
section, we will also present the numerical results for the Wilson coefficients,
while the analytical expressions of these coefficients are collected in the 
appendix.

\begin{figure}
\begin{center}
\epsfxsize=14cm
\epsfbox{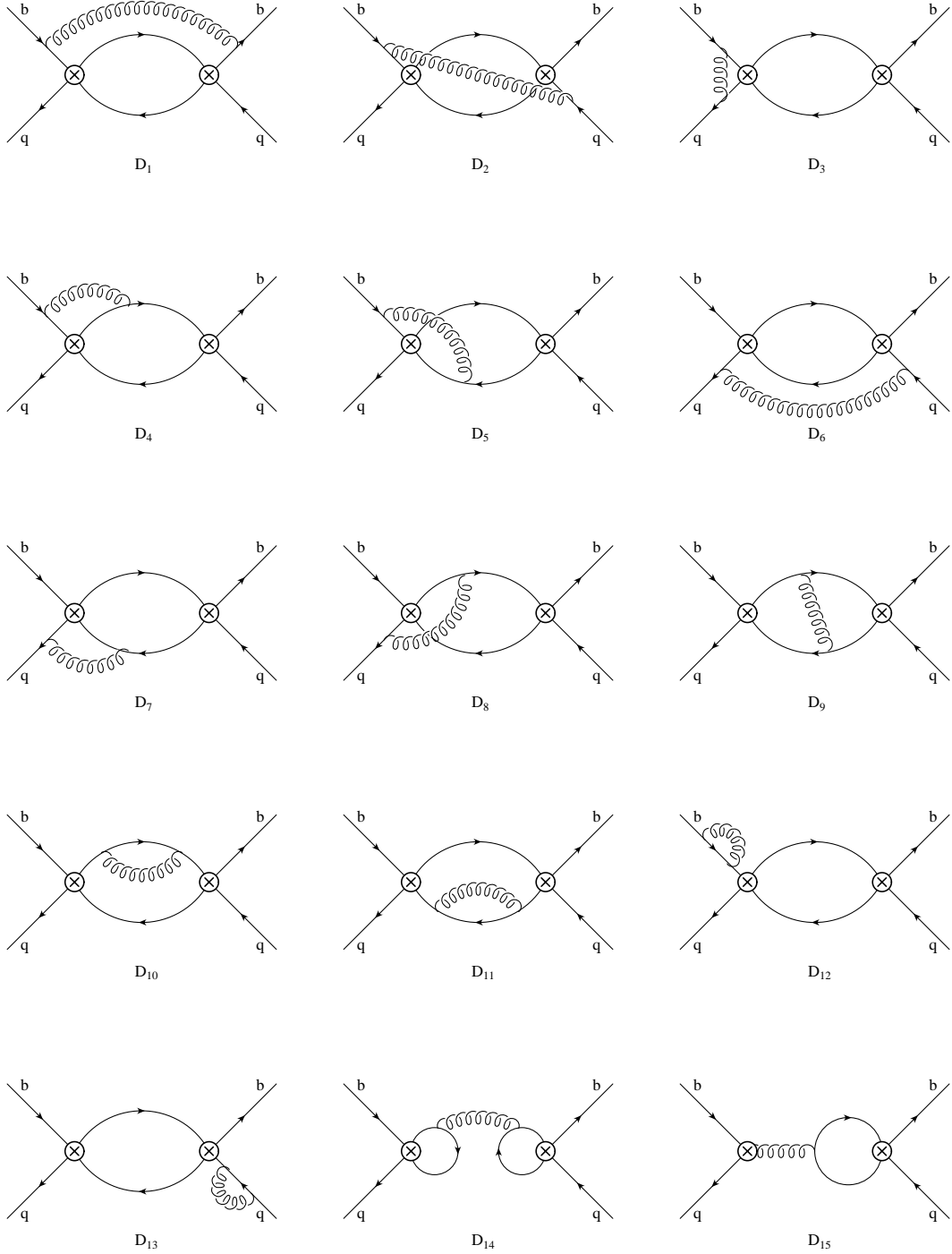}
\end{center}
\caption{\it Feynman diagrams which contribute at NLO to the matrix element
of the transition operator ${\cal T}$ in the case $q=s$. In the other cases, 
$q=u,\,d$, diagrams $D_{14}$ and $D_{15}$ are Cabibbo suppressed and have been 
neglected in the calculation.}
\label{fig:fddb1}
\end{figure}
\begin{figure} [t]
\begin{center}
\epsfxsize=12cm
\epsfbox{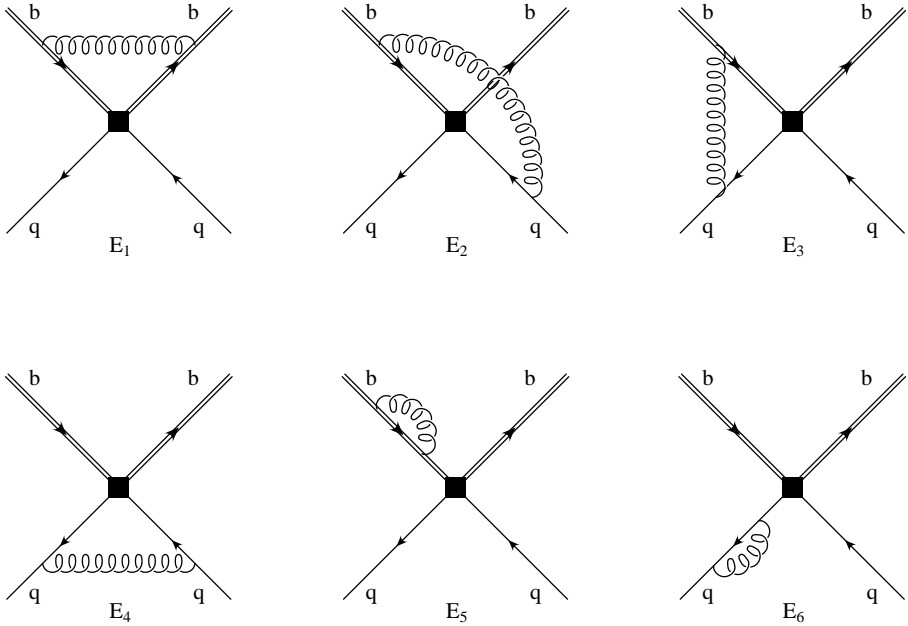}
\end{center}
\caption{\it Feynman diagrams which contribute, at NLO, to the matrix
element of the $\DB=0$ operators entering the HQET.}
\label{fig:diaeff}
\end{figure}

In order to compute the Wilson coefficients of the $\DB=0$ operators at the 
NLO, we have evaluated in QCD the imaginary part of the diagrams shown in 
fig.~\ref{fig:fddb1} ({\it full} theory) and in the HQET the diagrams shown in 
fig.~\ref{fig:diaeff} ({\it effective} theory). The external quark states have 
been taken on-shell and all quark masses, except $m_b$ and $m_c$, have been 
neglected. More specifically, we have chosen the heavy quark momenta $p_b^2=
m_b^2$ in QCD and $k_b=0$ in the HQET, and $p_q=0$ for the external light 
quarks. In this way, we automatically retain the leading term in the $1/m_b$ 
expansion. We have performed the calculation in a generic covariant gauge, in 
order to check the gauge independence of the final results. Two-loop integrals 
have been reduced to a set of independent master integrals by using the 
recurrence relation technique~\cite{integrals1}-\cite{integrals3} implemented 
in the TARCER package~\cite{tarcer}. Equations of motion have been used to 
reduce the number of independent operators.

Some diagrams, both in the full and in the effective theory, are plagued by IR 
divergences. These divergences do not cancel in the final partonic amplitudes, 
and provide an example of violation of the Bloch-Nordsieck theorem in 
non-abelian gauge theories~\cite{Bloch:1937pw}-\cite{Di'Lieto:1980dt}. 
IR poles, however, are expected to cancel in the matching and we have 
explicitly checked that, in the computation of the coefficient functions, this 
cancellation takes place.

We use $D$-dimensional regularization with anticommuting $\gamma_5$ (NDR) to
regularize both UV and IR divergences. As discussed in details in
ref.~\cite{noantri}, the presence of dimensionally-regularized IR divergences 
introduces subtleties in the matching procedure. The matching must be 
consistently performed in $D$ dimensions. This requires, in particular, 
enlarging the operator basis to include (renormalized) evanescent operators, 
which must be inserted in the one-loop diagrams of the effective theory. 
Because of the IR divergences, the matrix elements of the renormalized 
evanescent operators do not vanish in the $D\to 4$ limit~\cite{Misiak:1999yg},
and give a finite contribution in the matching procedure.

As a check of the perturbative calculation, we have verified that our results
for the Wilson coefficients satisfy the following requirements:
\begin{itemize}
\item {\it gauge invariance}:
the coefficient functions in the $\overline{\rm{MS}}$ scheme are explicitly 
gauge-invariant. The same is true for the full and the effective amplitudes 
separately;
\item {\it renormalization-scale dependence}:
the coefficient functions have the correct logarithmic scale dependence as 
predicted by the LO anomalous dimensions of the $\DB=0$ and $\DB=1$ operators;
\item {\it IR divergences}: 
the coefficient functions are infrared finite. We verified that the 
cancellation of IR divergences also takes place, separately in the full and 
the effective amplitudes, for the abelian combination of diagrams .
\end{itemize}

The $\DB=0$ effective theory is derived from the double insertion of the 
$\DB=1$ effective hamiltonian. Therefore, the coefficient functions $c^q_k$ of 
the $\DB=0$ effective theory depend quadratically on the coefficient functions 
$C_i$ of the $\DB=1$ effective hamiltonian, and we can write
\be
c^q_k(\mu_0) = \sum_{i,j} C_i(\mu_1)\, C_j(\mu_1)\,
F^q_{k,ij}(\mu_1,\mu_0)\,.
\label{eq:c6db0}
\ee
Eq.~(\ref{eq:c6db0}) shows explicitly the scale dependence of the several terms.
We denote with $\mu_1$ the renormalization scale of the $\DB=1$ effective 
hamiltonian, whereas $\mu_0$ is the renormalization scale of the $\DB=0$ 
operators. The coefficients $F^q_{k,ij}$ depend on the renormalization scheme 
and scale of both the $\DB=0$ and $\DB=1$ operators. The dependence on the scale
$\mu_1$ and on the renormalization scheme of the $\DB=1$ operators actually 
cancels, order by order in perturbation theory, against the corresponding 
dependence of the $\DB=1$ Wilson coefficients $C_i$. Therefore, the coefficient 
functions $c^q_k$ only depend on the renormalization scheme and scale of the 
$\DB=0$ operators. We have chosen to renormalize these operators, in the HQET, 
in the $\msb$ scheme defined in details in ref.~\cite{reyes}.

The four fermion operators, in the effective $\DB=1$ hamiltonian, are naturally
expressed in terms of the weak eigenstates $d^{\, \prime}$ and $s^{\, \prime}$.
For this reason, we can write the coefficients $F^q_{k,ij}$ of 
eq.~(\ref{eq:c6db0}) in the form
\bea
&& F^d_{k,ij} = \cos^2 \theta_c F^{d^{\, \prime}}_{k,ij} + 
               \sin^2 \theta_c F^{s^{\, \prime}}_{k,ij} \nn \\
&& F^s_{k,ij} = \sin^2 \theta_c F^{d^{\, \prime}}_{k,ij} + 
               \cos^2 \theta_c F^{s^{\, \prime}}_{k,ij}
\eea
for $i,j=1,2$. In the case $q=s$, the coefficient functions also receive 
contributions from the insertion of the penguin and chromomagnetic operators 
(diagram $D_{15}$ of fig.~\ref{fig:fddb1}). Since the Wilson coefficients 
$C_3$--$C_6$ are small, contributions with a double insertion of penguin 
operators can be safely neglected. As suggested in \cite{Beneke:1999sy}, a 
consistent way for implementing this approximation is to consider the 
coefficients $C_3$--$C_6$ as formally of ${\cal O}(\alpha_s)$. Within this 
approximation, only single insertions of penguin operators need to be 
considered at the NLO, and we can write
\be
\label{eq:c6sdb0}
c^s_k = \dsum_{i,j=1,2} C_i \, C_j F^s_{k,ij}
+ 2 \,\frac{\as}{4\pi} \, C_2 \, C_{8G} P_{k,28}
+ 2\, \dsum_{i=1,2} \dsum_{r=3,6} C_i \, C_r P_{k,ir}\,,
\ee
which generalizes eq.~(\ref{eq:c6db0}) for the case $q=s$.

\begin{table} [t]
\begin{center}
\begin{tabular}{||l||c|c|c||c|c|c||c|c|c||}
\cline{2-10}
\multicolumn{1}{c||}{} & \multicolumn{3}{c||}{$q=d$} &
\multicolumn{3}{c||}{$q=u$} & \multicolumn{3}{c||}{$q=s$} \\
\cline{2-10} 
\multicolumn{1}{c||}{} & LO & NLO & NLO &  LO & NLO  & NLO & LO & NLO  & NLO \\
\multicolumn{1}{c||}{} &    &  $(m_c=0)$ &    &     & $(m_c=0)$ &    &    & $(m_c=0)$ &  \\ \hline \hline 
$ c^{\,q}_1$ & $-0.02 $ & $-0.03 $& $-0.03$ & $ -0.06$ & $-0.33 $ & $-0.29$& $ -0.01$ & $-0.03 $& $-0.03$ \\
$ c^{\,q}_2$ & $ 0.02$  & $0.03$ & $0.03$   & $0.00 $  & $-0.01 $& $-0.01$ & $0.02 $   & $0.03 $ & $0.04$ \\
$ c^{\,q}_3$ & $-0.70 $ & $-0.65 $ & $-0.67 $ & $2.11 $  & $2.27 $   & $2.34 $ & $-0.61 $  & $-0.51 $ & $-0.56$  \\
$ c^{\,q}_4$ & $0.79 $  & $0.68$  & $0.68$   & $0.00 $  & $-0.06 $   & $-0.05 $ & $0.77 $  & $0.61 $  & $0.64 $\\
\hline
\end{tabular}
\end{center}
\caption{\it 
Wilson coefficients $c^q_k$ computed at the LO and NLO, in the latter case with 
and without the inclusion of charm quark mass corrections at ${\cal O}(\as)$. 
As reference values of the input parameters, we use $\mu_0=\mu_1=m_b=4.8$ GeV 
and $m_c=1.4$ GeV.}
\label{tab:coeff}
\end{table}
The analytical expressions of the coefficients $F^q_{k,ij}$ and 
$P_{k,ij}$ will be given in the appendix. For illustration we present here, in 
table \ref{tab:coeff}, the numerical values of the coefficients $c_k^q$ both at 
LO and NLO and, in the latter case, with and without the inclusion of the
${\cal O}(\as)$ charm quark mass corrections computed in this paper. As 
reference values of the input parameters, we use $\mu_0=\mu_1=m_b=4.8$ GeV and 
$m_c=1.4$ GeV. 

By looking at the results shown in table \ref{tab:coeff}, we see that the NLO 
charm quark mass corrections are rather large for some of the Wilson 
coefficients, although the total effect of these contributions on the lifetime 
ratios will be found to be small. The numerical expressions of the lifetime 
ratios as a function of the $B$-parameters will be given in the next section 
(eq.(\ref{eq:magic})). In these expressions we will also include an estimate of 
the theoretical error on the coefficients, coming from the residual NNLO 
dependence on the renormalization scale $\mu_1$ and from the uncertainties on 
the values of the charm and bottom quark masses and the other input parameters.

The combinations $c_k^u-c_k^d$ ($k=1,\ldots 4$) of Wilson coefficients, which 
enter the theoretical expression of the ratio $\rp$, have been also computed,
at the NLO, in ref.~\cite{bucha}. In this case, the corresponding operators are
the flavour non-singlet combinations ${\cal O}_k^u-{\cal O}_k^d$, which do not 
mix, with coefficients proportional to powers of the $b$ quark mass, with 
operators of lower dimension. For this reason, the HQE can be also expressed in 
this case in terms of operators defined in QCD, and this is the choice followed 
by ref.~\cite{bucha}. 

In order to compare our results with those of ref.~\cite{bucha}, at the NLO, it 
is necessary to implement the matching, at ${\cal O}(\as)$, between QCD and HQET
operators. This matching can be written in terms of the Wilson coefficients, in 
the form
\be 
\label{eq:match}
\left[c_k^{\,u}(m_b)-c_k^{\,d}(m_b)\right]_{\rm QCD} = 
\left( 1 + \frac{\as(m_b)}{4\pi} \, \widehat S \right)_{k,l}\, 
\left[c_l^{\,u}(m_b)-c_l^{\,d}(m_b)\right]_{\rm HQET}\, ,
\ee
where a common renormalization scale $\mu=m_b$ has been chosen for all the 
coefficients. The matrix $\widehat S$ has been computed in ref.~\cite{noantri}. 
It depends on the renormalization schemes of both QCD and HQET operators. In
this paper, we have chosen to renormalize the HQET operators in the $\msb$ 
scheme of ref.~\cite{reyes}. By choosing for the QCD operators the $\msb$ 
scheme of ref.~\cite{bucha}, defined in ref.~\cite{db2nlo1}, one finds that the 
matrix $\widehat S$ is given by~\footnote{This matrix differs from the matrix 
$\widehat C_1$ given in ref.~\cite{noantri} by a linear transformation, 
since a basis of operators different from eq.~(\ref{eq:opeff}) has been chosen 
in that paper. More specifically, $\widehat S = - M \widehat C_1^T M^{-1}$, 
where $M$ is defined in ref.~\cite{noantri}.}
\renewcommand{\arraystretch}{1.2}
\be
\label{eq:c1ms}
\widehat S=
\left(\begin{array}{cccc}
  32/3 & 0      & 22/9 & 1/3  \cr
 -16/3 & - 16/3 &  4/9 & -2/9 \cr
    11 & 3/2    & -7/2 &  1/4 \cr
    2  &  -1    & 3    &  5/2 \cr
\end{array}\right)
\ee
\renewcommand{\arraystretch}{2.0}
By using eq.~(\ref{eq:match}), we have verified that our results for the 
combinations $c_k^u-c_k^d$ agree with those of ref.~\cite{bucha}. Note that in
the notation of \cite{bucha} the labels $u$ and $d$ are interchanged with
respect to our convention and that the Wilson coefficients are defined with a
relative factor of 3. The numerical comparison of our results with those shown
in table 1 of ref.~\cite{bucha}, however, shows some differences, particularly
at the LO. The reason is that some of the coefficient functions, because of
large cancellations, are extremely sensitive to the value of the coupling 
constant $\as(\mu)$.

\section{Theoretical estimates of the lifetime ratios}
\label{sec:phenom}

In this section we present the theoretical estimates of the lifetime ratios of 
beauty hadrons, obtained by using the NLO expressions of the Wilson 
coefficients and the lattice determinations of the relevant hadronic matrix
elements~\cite{DiPierro98}-\cite{APE}.

The HQE for the ratio of inclusive widths of beauty hadrons is expressed by
eq.~(\ref{eq:ratio}), up to and including $1/m_b^3$ spectator effects. The 
combinations of hadronic parameters entering this formula at order $1/m_b^2$ 
can be evaluated from the heavy hadron spectroscopy~\cite{mupig}, and one 
obtains the estimate
\be
\label{eq:del}
\dfrac{\tau(B^+)}{\tau(B_d)}= 1.00 -\Delta^{B^+}_{\rm\scriptstyle spec}\,,\quad
\dfrac{\tau(B_s)}{\tau(B_d)}= 1.00 -\Delta^{B_s}_{\rm\scriptstyle spec}\,,\quad
\dfrac{\tau(\Lambda_b)}{\tau(B_d)}= 0.98(1)-
\Delta^{\Lambda}_{\rm\scriptstyle spec} \,,
\ee
where the $\Delta$s represent the $1/m_b^3$ contributions of hard spectator 
effects
\be
\label{eq:delta}
\Delta^{H_b}_{\rm\scriptstyle spec}= 
\dfrac{96\pi^2}{m_b^3\, c^{(3)}}\dsum\limits_{k} c^{(6)}_k \left(\frac{\langle 
O^{(6)}_k\rangle_{H_b}}{2M_{H_b}}-\frac{\langle O_k^{(6)}\rangle_{B_d}}
{2M_{B_d}}\right) \, .
\ee
Note that, by neglecting these contributions, the $1/m_b^2$ predictions of
eq.~(\ref{eq:del}) are incompatible with the experimental results of
eq.~(\ref{eq:rexp}).

In parametrizing the matrix elements of the dimension-six current-current 
operators, we follow the analysis of ref.~\cite{noantri} and we distinguish two 
cases, depending on whether or not the light quark $q$ of the operator enters 
as a valence quark in the external hadronic state. Correspondingly, we 
introduce different $B$-parameters for the valence and non-valence 
contributions. For the $B$-meson matrix elements, we write the matrix elements 
of the non-valence operators in the form
\be
\dfrac{\langle B_{q}\vert {\cal O}^{q'}_k\vert B_{q}\rangle}{2M_{B_q}}=
\dfrac{f_{B_q}^2M_{B_q}}{2} \, \delta_k^{\,q'q} \quad {\rm for} \,\, q \neq q'
\,,
\label{eq:bpar1}
\ee
while, in the case of the valence contributions ($q=q'$), we write
\be
\begin{array}{ll}
\dfrac{\langle B_q\vert {\cal O}^q_1 \vert B_q\rangle}{2M_{B_q}}=
\dfrac{f_{B_q}^2M_{B_q}}{2} \, \left( B_1^{\, q} +
\delta_1^{\, qq}\right) \, , \quad
\dfrac{\langle B_q\vert {\cal O}^q_3 \vert B_q\rangle}{2M_{B_q}}=
\dfrac{f_{B_q}^2M_{B_q}}{2} \, \left( \ep_1^{\, q} +
\delta_3^{\, qq}\right) \, , \\
\dfrac{\langle B_q\vert {\cal O}^q_2 \vert B_q\rangle}{2M_{B_q}}=
\dfrac{f_{B_q}^2M_{B_q}}{2} \, \left( B_2^{\, q} +
\delta_2^{\, qq}\right) \, , \quad
\dfrac{\langle B_q\vert {\cal O}^q_4 \vert B_q\rangle}{2M_{B_q}}=
\dfrac{f_{B_q}^2M_{B_q}}{2} \, \left( \ep_2^{\, q} +
\delta_4^{\, qq}\right) \,.
\end{array}
\label{eq:bpar2}
\ee

The parameters $\delta_k^{\, qq}$ in eq.~(\ref{eq:bpar2}) are defined as the
$\delta_k^{\, qq'}$ of eq.~(\ref{eq:bpar1}) in the limit of degenerate quark
masses ($m_q=m_{q'}$). In the VIA, $B_1^{\, q}=B_2^{\, q}=1$ while the $\ep$
parameters and all the $\delta$s vanish. Note that, in the $SU(2)$ limit, the 
parameters $B_{1,2}^{\, d}$ and $\ep_{1,2}^{\, d}$ express the matrix elements 
of the non-singlet operator ${\cal O}^u_k-{\cal O}^d_k$ between external 
$B$-meson states.

The reason to distinguish between valence and non valence contributions, is 
that only the former have been computed so far by using lattice QCD 
simulations~\cite{DiPierro98,APE}. A non-perturbative lattice calculation of 
the $\delta$ parameters would be also possible, in principle. However, it 
requires to deal with the difficult problem of subtractions of
power-divergences, which has prevented so far the calculation of the 
corresponding diagrams.

To complete the definitions of the $B$-parameters for the $B$-mesons, we
introduce a parameter for the matrix element of the penguin operator
\be
\dfrac{\langle B_q\vert {\cal O}_P\vert B_q\rangle}{2M_{B_q}}=
\dfrac{f_{B}^2M_{B}}{2} \, P^{\, q} \,.
\label{eq:bpar3}
\ee

We now define the $B$-parameters for the $\Lambda_b$ baryon. Up to $1/m_b$
corrections, the matrix elements of the operators ${\cal O}^q_2$ and ${\cal
O}^q_4$, between external $\Lambda_b$ states, can be related to the matrix
elements of the operators ${\cal O}^q_1$ and ${\cal O}^q_3$~\cite{NS}
\be
\langle \Lambda_b\vert {\cal O}^q_1 \vert \Lambda_b\rangle = -2 \,
\langle \Lambda_b\vert {\cal O}^q_2 \vert \Lambda_b\rangle \, , \quad
\langle \Lambda_b\vert {\cal O}^q_3 \vert \Lambda_b\rangle = -2 \,
\langle \Lambda_b\vert {\cal O}^q_4 \vert \Lambda_b\rangle \, .
\ee
For the independent matrix elements, assuming $SU(2)$ symmetry, we define
\bea
\label{eq:bpar4}
&& \dfrac{\langle \Lambda_b\vert {\cal O}^q_1 \vert \Lambda_b\rangle}
{2M_{\Lambda_b}}= \dfrac{f_{B}^2M_{B}}{2} \, \left( L_1 +
\delta_1^{\,\Lambda q}\right) \quad {\rm for} \,\, q=u,d \, ,\nonumber \\
&& \dfrac{\langle \Lambda_b\vert {\cal O}^q_3 \vert \Lambda_b\rangle}
{2M_{\Lambda_b}}= \dfrac{f_{B}^2M_{B}}{2} \, \left( L_2 +
\delta_2^{\,\Lambda q}\right) \quad {\rm for} \,\, q=u,d  \, , \nonumber \\
&& \dfrac{\langle \Lambda_b\vert {\cal O}^q_1\vert \Lambda_b\rangle}
{2M_{\Lambda_b}}= \dfrac{f_{B}^2M_{B}}{2} \, \delta_1^{\, \Lambda q}
\quad {\rm for}\,\, q=s,c   \, ,\\
&& \dfrac{\langle \Lambda_b\vert {\cal O}^q_3\vert \Lambda_b\rangle}
{2M_{\Lambda_b}}= \dfrac{f_{B}^2M_{B}}{2} \, \delta_2^{\, \Lambda q}
\quad {\rm for}\,\, q=s,c   \, ,\nonumber \\
&& \dfrac{\langle \Lambda_b\vert {\cal O}_P\vert \Lambda_b\rangle}
{2M_{\Lambda_b}}=\dfrac{f_{B}^2M_{B}}{2} \, P^{\, \Lambda} \nonumber \,.
\eea
In analogy with the $B$-meson case, the parameters $L_1$ and $L_2$ represent
the valence contributions computed so far by current lattice
calculations~\cite{DiPierro:1999tb,DiPierroproc}.

In terms of valence and non valence $B$-parameters, the quantities $\Delta^{H_b}
_{\rm\scriptstyle spec}$ of eq.~(\ref{eq:delta}), which represent the spectator 
contributions to the lifetime ratios, are expressed in the form
\bea
\label{eq:masterformula}
\Delta^{B^+}_{\rm\scriptstyle spec}&=& 48\pi^2\, \frac{f_B^2 M_B}{m_b^3 c^{(3)}}
\, \dsum_{k=1}^4 \left( c_k^{\,u}- c_k^{\,d} \right)
{\cal B}_k^{\, d} \, , \nn \\
\Delta^{B_s}_{\rm\scriptstyle spec}&=&
48\pi^2\, \frac{f_B^2 M_B}{m_b^3 c^{(3)}} \, \left\{ \dsum_{k=1}^4
\left[r \,  c_k^{\,s} \, {\cal B}_k^{\, s} -  c_k^{\,d} \,
{\cal B}_k
^{\, d} + \left(  c_k^{\,u} +  c_k^{\,d} \right) \left(r\,
\delta_k^{\,ds} - \delta_k^{\,dd} \right) + \right. \right.  \nn \\ && \qquad
\qquad \quad \left. \left.
 c_k^{\,s} \left(r \, \delta_k^{\,ss} - \delta_k^{\,sd} \right) +
 c_k^{\,c}
\left(r \, \delta_k^{\,cs} - \delta_k^{\,cd} \right) \right] + c_P
\left(r P^{\, s} - P^{\, d} \right) \right\} \, ,\\
\Delta^{\Lambda}_{\rm\scriptstyle spec}&=&
48\pi^2\, \frac{f_B^2 M_B}{m_b^3 c^{(3)}} \, \left\{ \dsum_{k=1}^4 \left[\left(
  c_k^{\,u} +  c_k^{\,d} \right) {\cal L}_k^{\, \Lambda} -
  c_k^{\,d} \, {\cal B}_k^{\, d} + \left(  c_k^{\,u} +
 c_k^{\,d} \right) \left(\delta_k^{\,\Lambda d} - \delta_k^{\,dd} \right)
+ \right. \right.  \nn \\
&& \qquad \qquad \quad \left. \left.  c_k^{\,s}\left(\delta_k^{\,
\Lambda s}- \delta_k^{\,sd} \right) +  c_k^{\,c} \left(\delta_k^{\,
\Lambda c}- \delta_k^{\,cd} \right) \right] + c_P \left(P^{\,
\Lambda} - P^{\, d} \right)
\right\} \, . \nn
\eea
where $r$ denotes the ratio $(f_{B_s}^2 M_{B_s})/(f_B^2 M_B)$ and, in order to 
simplify the notation, we have defined the vectors of parameters
\bea
&& \vec {\cal B}^q=\{B_1^q,B_2^q,\ep_1^q,\ep_1^q\} \, ,\nonumber \\
&& \vec {\cal L}=\{L_1,-L_1/2,L_2,-L_2/2\} \, ,\\
&& \vec \delta^{\Lambda q}=\{\delta^{\Lambda q}_1,-\delta^{\Lambda q}_1/2,
\delta^{\Lambda q}_2,-\delta^{\Lambda q}_2/2\} \nonumber \, .
\eea

An important consequence of eq.~(\ref{eq:masterformula}) is that, because of 
the $SU(2)$ symmetry, the non-valence ($\delta$s) and penguin ($P$s)
contributions cancel out in the expressions of the lifetime ratio $\rp$. Thus,
the theoretical prediction of this ratio is at present the most accurate, since 
it depends only on the non-perturbative parameters actually computed by current 
lattice calculations. The prediction of the ratio $\rl$, instead, is affected 
by both the uncertainties on the values of the $\delta$ and $P$ parameters, and 
by the unknown expressions of the Wilson coefficients $c_k^{\, c}$ and $c_P$ at 
the NLO. For the ratio $\rs$ the same uncertainties exist, although their effect
is expected to be smaller, since the contributions of non-valence and penguin
operators cancel, in this case, in the limit of exact $SU(3)$ symmetry.

In the numerical analysis of the ratios $\rs$ and $\rl$, we will neglect the 
non-valence and penguin contributions. The non-valence contributions vanish in
the VIA, and present phenomenological estimates indicate that the corresponding 
matrix elements are suppressed, with respect to the valence contributions, by 
at least one order of magnitude~\cite{Chernyak:1995cx,pirjol}. On the other 
hand, the matrix elements of the penguin operators are not expected to be 
smaller than those of the valence operators. Since the coefficient function 
$c_P$ vanishes at the LO, this contribution is expected to have the size of a 
typical NLO corrections. Thus, from a theoretical point of view, a quantitative 
evaluation of the non-valence and penguin operator matrix elements would be of 
the greatest interest to improve the determination of the $\Lambda_B$ lifetime.

By neglecting the non valence and penguin contributions, we obtain from 
eq.~(\ref{eq:masterformula}) the following numerical expressions for the
spectator effects
\be
\label{eq:magic}
\renewcommand{\arraystretch}{1.2}
\begin{array}{ll}
\Delta^{B^+}_{\rm\scriptstyle spec} \, =&
-\, 0.06(2) \, B_1^d - 0.010(3) \, B_2^d + 0.7(2) \, \ep_1^d -0.18(5) \, \ep_2^d
\,,\\ \\
\Delta^{B_s}_{\rm\scriptstyle spec} \, =&
-\, 0.010(2) \, B_1^s + 0.011(3) \, B_2^s - 0.16(4)\,\ep_1^s + 0.18(5)\,\ep_2^s
\\ &
+\, 0.008(2)\, B_1^d - 0.008(2) \, B_2^d + 0.16(4) \, \ep_1^d -0.16(4)\, \ep_2^d
\,,\\ \\
\Delta^{\Lambda}_{\rm\scriptstyle spec} \, =&
-\, 0.08(2)\, L_1 + 0.33(8)\, L_2 \\ &
+\, 0.008(2)\, B_1^d - 0.008(2) \, B_2^d + 0.16(4) \, \ep_1^d -0.16(4)\, \ep_2^d
\,,\end{array}
\renewcommand{\arraystretch}{2.0}
\ee
These formulae, which are accurate at the NLO, represent the main result of this
paper. The errors on the coefficients take into account both the residual NNLO 
dependence on the renormalization scale of the $\DB=1$ operators and the
theoretical uncertainties on the input parameters. To estimate the former, the 
scale $\mu_1$ has been varied in the interval between $m_b/2$ and $2 m_b$. For 
the charm and bottom quark masses, and the B meson decay constants we have used 
the central values and errors given in table \ref{tab:inputs}. The strong 
coupling constant has been fixed at the value $\as(m_Z)=0.118$. The parameter 
$c^{(3)}$ in eq.~(\ref{eq:masterformula}) is a function of the ratio $m_c^2/
m_b^2$, and such a dependence has been consistently taken into account in the 
numerical analysis and in the estimates of the errors. For the range of masses 
given in table \ref{tab:inputs}, $c^{(3)}$ varies in the interval $c^{(3)}=3.4
\div 4.2$~\cite{gamma4}-\cite{gamma6}.
\begin{table} [t]
\renewcommand{\arraystretch}{1.6}
\begin{center}
\begin{tabular}{|cc|}
\hline
$ B_1^d$ = $1.2 \pm 0.2$ & $ B_1^s$ = $1.0 \pm 0.2$  \\
$ B_2^d$ = $0.9 \pm 0.1$ & $ B_2^s$ = $0.8 \pm 0.1$ \\
$ \ep_1^d$ = $0.04 \pm 0.01$ & $ \ep_1^s$ = $0.03 \pm 0.01$ \\
$ \ep_2^d$ = $0.04 \pm 0.01$ & $ \ep_2^s$ = $0.03 \pm 0.01$ \\
\hline
$ L_1$ = $-0.2 \pm 0.1$ & $ L_2$ = $0.2 \pm 0.1$ \\
\hline
$m_b$ = $4.8\pm 0.1$ GeV & $m_b-m_c$ = $3.40\pm 0.06$ GeV \\
$f_B$ = $200\pm 25$ MeV & $f_{B_s}/f_B$ = $1.16\pm 0.04$ \\
\hline
\end{tabular}
\end{center}
\caption{\it Central values and standard deviations of the input parameters used
in the numerical analysis. The values of $m_b$ and $m_c$ refer to the pole mass
definitions of these quantities.}
\label{tab:inputs}
\end{table}

As discussed in the previous section, the HQE for the ratio $\rp$ can be also
expressed in terms of operators defined in QCD. The corresponding coefficient
functions can be evaluated by applying the matching defined in 
eq.~(\ref{eq:match}). In this way, we obtain the expression
\be
\label{eq:magic_qcd}
\Delta^{B^+}_{\rm\scriptstyle spec} =
-\, 0.05(1) \, \bar B_1^d - 0.007(2) \, \bar B_2^d + 0.7(2) \, \bar \ep_1^d 
-0.15(4) \, \bar \ep_2^d
\ee
where the $\bar B$ and $\bar \ep$ parameters are now defined in terms of matrix 
elements of QCD operators. This expression is in agreement with the result 
obtained in ref.~\cite{bucha}.

The errors quoted on the coefficients in eq.~(\ref{eq:magic}) are strongly 
correlated, since they originate from the theoretical uncertainties on the same 
set of input parameters. For this reason, in order to evaluate the lifetime 
ratios, we have not used directly eq.~(\ref{eq:magic}). Instead, we have 
performed a bayesian statistical analysis by implementing a short Monte Carlo 
calculation. The input parameters have been extracted with flat distributions,
assuming as central values and standard deviations the values given in table 
\ref{tab:inputs}. The results for the $B$-parameters are based on the lattice 
determinations of refs.~\cite{DiPierro98}-\cite{APE}.\footnote{For recent 
estimates of these matrix elements based on QCD sum rules, see 
refs.~\cite{Colangelo:1996ta}-\cite{Huang:1999xj}.} As discussed in details 
in ref.~\cite{noantri}, we have included in the errors an estimate of the 
uncertainties not taken into account in the original papers. The QCD results 
for the $B$ meson $B$-parameters of ref.~\cite{APE} have been converted to HQET 
by using eq.~(\ref{eq:match}).\footnote{With respect to ref.~\cite{noantri}, we
use for the $B$-meson $B$-parameters the results updated in ref.~\cite{APE}.}
The contributions of all the $\delta$ and $P$ parameters have been neglected.
\begin{figure}
\begin{center}
\epsfxsize=14cm
\epsfbox{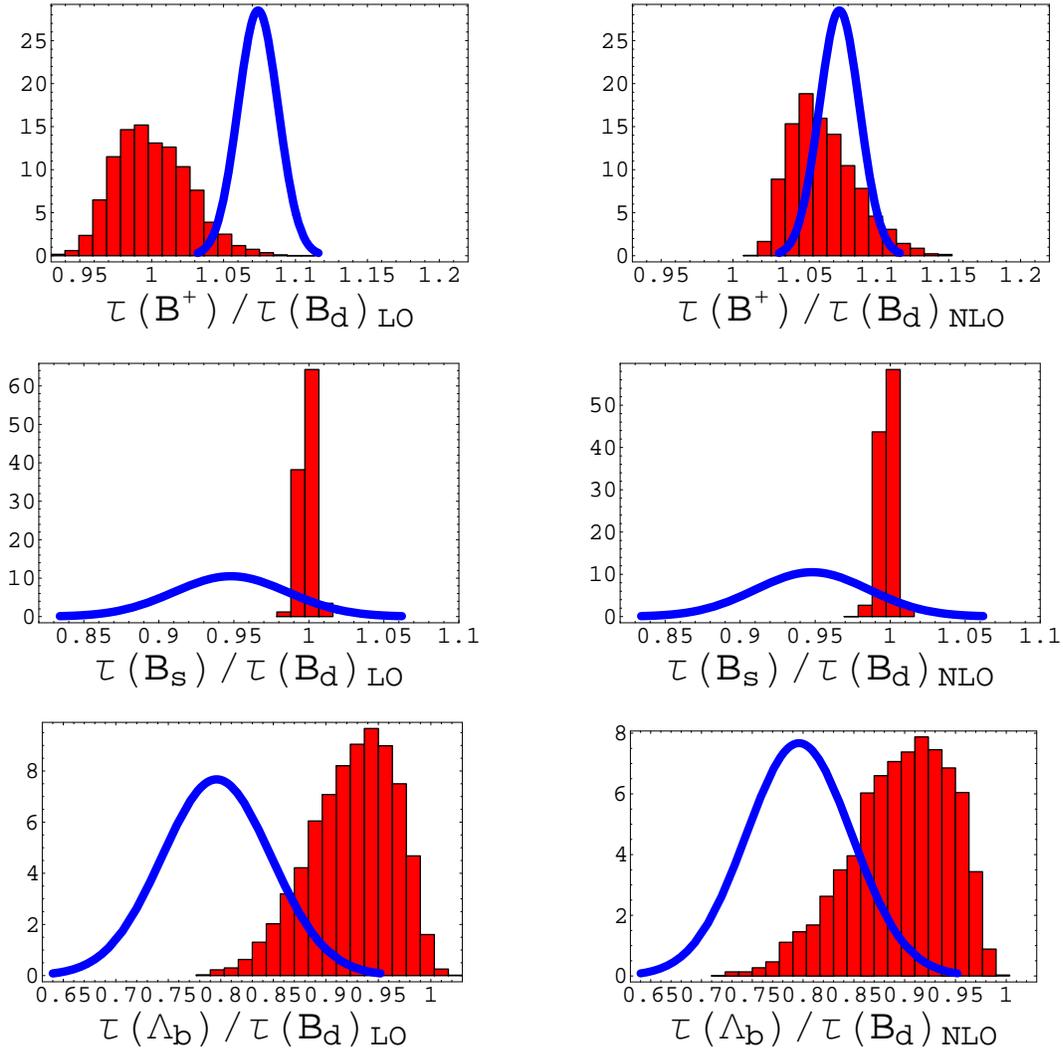}
\end{center}
\caption{\it Theoretical (histogram) vs experimental (solid line) distributions 
of lifetime ratios. The theoretical predictions are shown at the LO (left) and 
NLO (right).}
\label{fig:plot}
\end{figure}

In this way we obtain the NLO predictions for the lifetimes ratios which have 
been also quoted in the introduction
\be
\label{eq:nlores2}
\frac{\tau(B^+)}{\tau(B_d)} = 1.06 \pm 0.02 \, ,  \qquad
\frac{\tau(B_s)}{\tau(B_d)} = 1.00 \pm 0.01 \, ,  \qquad
\frac{\tau(\Lambda_b)}{\tau(B_d)} = 0.90 \pm 0.05\, .
\ee
The central values and errors correspond to the average and the standard 
deviation of the theoretical distributions. These distributions are shown in 
fig.~\ref{fig:plot}, together with the experimental ones. We mention that
uncertainties coming from the residual scale dependence in the results of 
eq.~(\ref{eq:nlores2}) represent less than 20\% of the quoted errors. 

In conclusion we find that, with the inclusion of the NLO corrections, the 
theoretical prediction for the ratio $\rp$ turns out to be in very good 
agreement with the experimental measurement, given in eq.~(\ref{eq:rexp}). For
the ratios $\rs$ and $\rl$ the agreement is also very satisfactory, and the 
difference between theoretical and experimental determinations is at the 
$1\sigma$ level. We have pointed out, however, that the theoretical predictions
are less accurate in these cases, since a reliable estimate of the 
contribution of the non-valence and penguin operators cannot be performed yet. 
We also found that the NLO charm quark mass corrections computed in this paper
are rather large for some of the Wilson coefficients. Nevertheless, the total 
effect of these contributions on the lifetime ratios is numerically small.

\section*{Acknowledgments}
We are very grateful to M. Ciuchini for interesting discussions on the subject 
of this paper. Work  partially supported by the European Community's Human 
Potential Programme under HPRN-CT-2000-00145 Hadrons/Lattice QCD.

\section*{Appendix}
In this appendix we collect the analytical expressions of the Wilson
coefficients, both at the LO and NLO. The LO coefficients have been computed 
in refs.~\cite{NS,Ura} and are reported here for completeness.

We distinguish the leading and next-to-leading contributions in the 
coefficients $F^q_{k,ij}$ by writing the expansion
\be
F^q_{k,ij} = A^{q}_{k,ij} + \frac{\as}{4\pi} B^{q}_{k,ij}\,,
\ee
$(q=u,d^\prime,s^\prime)$. Since, by definition, the coefficients $A^q_{k,ij}$ 
and $B^q_{k,ij}$ are symmetric in the indices $i$ and $j$, we will only present 
results for $i \le j$. 

The LO coefficients $A^{q}_{k,ij}$ read
\renewcommand{\arraystretch}{2.5} 
\begin{footnotesize}
\be
\begin{array}{lll}
  A^u_{1,11}={\dfrac{{{\left( 1 - z \right) }^2}}{3}}\,, \qquad
& A^u_{1,12}={{\left( 1 - z \right) }^2}\,,\qquad
& A^u_{1,22}={\dfrac{{{\left( 1 - z \right) }^2}}{3}}\,,\\
  A^u_{2,11}=0\,, \qquad
& A^u_{2,12}=0\,, \qquad
& A^u_{2,22}=0\,, \\
  A^u_{3,11}=2\,{{\left( 1 - z \right) }^2}\,, \qquad
& A^u_{3,12}=0\,, \qquad
& A^u_{3,22}=2\,{{\left( 1 - z \right) }^2}\,,\\
  A^u_{4,11}=0\,, \qquad
& A^u_{4,12}=0\,, \qquad
& A^u_{4,22}=0\,. 
\end{array}
\ee

\be
\begin{array}{lll}
  A^{d^{\, \prime}}_{1,11}={\dfrac{-{{\left( 1 - z \right) }^2}\,\left( 2 + z \right)}
   {2}}\,, \qquad
& A^{d^{\, \prime}}_{1,12}={\dfrac{-{{\left( 1 - z \right) }^2}\,\left( 2 + z \right)}
   {6}}\,,\qquad
& A^{d^{\, \prime}}_{1,22}={\dfrac{-{{\left( 1 - z \right) }^2}\,\left( 2 + z \right)}
   {18}}\,,\\
  A^{d^{\, \prime}}_{2,11}={{\left( 1 - z \right) }^2}\,\left( 1 + 2\,z \right)\,, \qquad
& A^{d^{\, \prime}}_{2,12}={\dfrac{{{\left( 1 - z \right) }^2}\,\left( 1 + 2\,z \right) }{3}}\,, \qquad
& A^{d^{\, \prime}}_{2,22}={\dfrac{{{\left( 1 - z \right) }^2}\,\left( 1 + 2\,z \right) }{9}}\,, \\
  A^{d^{\, \prime}}_{3,11}=0\,, \qquad
& A^{d^{\, \prime}}_{3,12}=0\,, \qquad
& A^{d^{\, \prime}}_{3,22}={\dfrac{-{{\left( 1 - z \right) }^2}\,\left( 2 + z \right)}
   {3}}\,,\\
  A^{d^{\, \prime}}_{4,11}=0\,, \qquad
& A^{d^{\, \prime}}_{4,12}=0\,, \qquad
& A^{d^{\, \prime}}_{4,22}={\dfrac{2\,{{\left( 1 - z \right) }^2}\,\left( 1 + 2\,z \right)}{3}}\,.
\end{array}
\ee

\be
\begin{array}{lll}
  A^{s^{\, \prime}}_{1,11}=-{\sqrt{1 - 4\,z}}\,\left( 1 - z \right)\,, \qquad
& A^{s^{\, \prime}}_{1,12}={\dfrac{-{\sqrt{1 - 4\,z}}\,\left( 1 - z \right)}{3}}\,,\qquad
& A^{s^{\, \prime}}_{1,22}={\dfrac{-{\sqrt{1 - 4\,z}}\,\left( 1 - z \right)}{9}}\,,\\
  A^{s^{\, \prime}}_{2,11}={\sqrt{1 - 4\,z}}\,\left( 1 + 2\,z \right)\,, \qquad
& A^{s^{\, \prime}}_{2,12}={\dfrac{{\sqrt{1 - 4\,z}}\,\left( 1 + 2\,z \right) }{3}}\,, \qquad
& A^{s^{\, \prime}}_{2,22}={\dfrac{{\sqrt{1 - 4\,z}}\,\left( 1 + 2\,z \right) }{9}}\,, \\
  A^{s^{\, \prime}}_{3,11}=0\,, \qquad
& A^{s^{\, \prime}}_{3,12}=0\,, \qquad
& A^{s^{\, \prime}}_{3,22}={\dfrac{-2\,{\sqrt{1 - 4\,z}}\,\left( 1 - z \right) }{3}}\,,\\
  A^{s^{\, \prime}}_{4,11}=0\,, \qquad
& A^{s^{\, \prime}}_{4,12}=0\,, \qquad
& A^{s^{\, \prime}}_{4,22}={\dfrac{2\,{\sqrt{1 - 4\,z}}\,\left( 1 + 2\,z \right) }{3}}\,. 
\end{array}
\ee
\end{footnotesize}
\renewcommand{\arraystretch}{2.0} 
where $z=m_c^2/m_b^2$.

The NLO results for the coefficients $B^{q}_{k,ij}$ have been obtained in the 
$\msb$ scheme of ref.~\cite{db2nlo1} for the $\DB=1$ operators and the $\msb$ 
scheme of ref.~\cite{reyes} for the $\DB=0$, HQET operators. We find
\renewcommand{\arraystretch}{2.5} 
\begin{footnotesize}
\bea
B^u_{1,11}&=&{\frac{-8\,\left( 1 - z \right) \,
      \left( 164 + 12\,{{\pi }^2} - 109\,z + 5\,{z^2} \right) }{81}} - 
  {\frac{16\,{{\left( 1 - z \right) }^2}\,
      \left( 6\,\log x_1 + 
        \left( 1 + z \right) \,\log (1 - z) \right) }{9}} - \nn\\
& &  {\frac{16\,z\,\left( 12 + z - 3\,{z^2} \right) \,\log z}{27}} - 
  {\frac{16\,\left( 1 - 3\,z \right) \,\left( 1 - z \right) \,
      \left( \log (1 - z)\,\log z + 2\,\mathrm{Li_2}(z) \right) }{9}
    }\,,\nn\\
B^u_{2,11}&=&{\frac{16\,\left( 1 - z \right) \,\left( 4 - 11\,z + 10\,{z^2} \right) }
    {81}} + {\frac{16\,{z^2}\,\log z}{27}}\,,\nn\\
B^u_{3,11}&=&{\frac{\left( \left( 1 - z \right) \,
        \left( 751 - 12\,{{\pi }^2}\,\left( 7 - 9\,z \right)  - 1199\,z - 
          26\,{z^2} \right)  \right) }{27}} +\nn\\ 
& &  8\,\log x_1 {{\left( 1 - z \right) }^2}\,- 
     {\frac{4\,\left( 26 - z \right)\,{{\left( 1 - z \right) }^2} \,\log (1 - z)}{3}}  - \nn\\
& &  {\frac{2\,z\,\left( 192 - 119\,z + 6\,{z^2} \right) \,\log z}{9}} - 
  {\frac{8\,\left( 4 - 3\,z \right) \,\left( 1 - z \right) \,
      \left( \log (1 - z)\,\log z + 2\,\mathrm{Li_2}(z) \right) }{3}
    }\,,\nn\\
B^u_{4,11}&=&{\frac{-8\,\left( 1 - z \right) \,\left( 11 - 10\,z - 13\,{z^2} \right) }
    {27}} + {\frac{32\,{z^2}\,\log z}{9}}\,,
\eea

\bea
B^u_{1,12}&=&{\frac{-8\,\left( 1 - z \right) \,
      \left( 13 + 2\,{{\pi }^2} - 2\,z + {z^2} \right) }{9}} - 
  {\frac{4\,{{\left( 1 - z \right) }^2}\,
      \left( 6\,\log x_0 + 4\,\log (1 - z) \right) }{3}} -\nn\\ 
& &  {\frac{16\,\left( 3 - z \right) \,z\,\log z}{3}} + 
  {\frac{16\,\left( 1 - z \right) \,z\,
      \left( \log (1 - z)\,\log z + 2\,\mathrm{Li_2}(z) \right) }{3}
    }\,,\nn\\
B^u_{2,12}&=&{\frac{32\,{{\left( 1 - z \right) }^3}}{9}}\,,\nn\\
B^u_{3,12}&=&{\frac{-2\,\left( 1 - z \right) \,
      \left( 125 + 6\,{{\pi }^2} - 103\,z + 2\,{z^2} \right) }{9}} +\nn\\ 
& &  {{\left( 1 - z \right) }^2}\,\left( 6\,\log x_0 - 
     24\,\log x_1 - 4\,z\,\log (1 - z) \right)  - 
  {\frac{4\,z\,\left( 3 + 4\,z - 3\,{z^2} \right) \,\log z}{3}} - \nn\\
& &  4\,\left( 1 - 2\,z \right) \,\left( 1 - z \right) \,
   \left( \log (1 - z)\,\log z + 2\,\mathrm{Li_2}(z) \right) \,,\nn\\
B^u_{4,12}&=&{\frac{-4\,\left( 1 - z \right) \,\left( 2 - z - 4\,{z^2} \right) }{9}} + 
  {\frac{4\,{z^2}\,\log z}{3}}\,,
\eea

\bea
B^u_{1,22}&=&{\frac{-8\,\left( 1 - z \right) \,
      \left( 164 + 12\,{{\pi }^2} - 109\,z + 5\,{z^2} \right) }{81}} -\nn\\ 
& &  {\frac{16\,{{\left( 1 - z \right) }^2}\,
      \left( 6\,\log x_1 + 
        \left( 1 + z \right) \,\log (1 - z) \right) }{9}} - 
  {\frac{16\,z\,\left( 12 + z - 3\,{z^2} \right) \,\log z}{27}} - \nn\\
& &  {\frac{16\,\left( 1 - 3\,z \right) \,\left( 1 - z \right) \,
      \left( \log (1 - z)\,\log z + 2\,\mathrm{Li_2}(z) \right) }{9}
    }\,,\nn\\
B^u_{2,22}&=&{\frac{16\,\left( 1 - z \right) \,\left( 4 - 11\,z + 10\,{z^2} \right) }
    {81}} + {\frac{16\,{z^2}\,\log z}{27}}\,,\nn\\
B^u_{3,22}&=&{\frac{\left( 1 - z \right) \,\left( 751 - 875\,z - 26\,{z^2} - 
        12\,{{\pi }^2}\,\left( 7 + 9\,z \right)  \right) }{27}} +\nn\\ 
& &  {\frac{4\,{{\left( 1 - z \right) }^2}\,
      \left( 6\,\log x_1 - 
        \left( 17 - z \right) \,\log (1 - z) \right) }{3}} - 
  {\frac{2\,\left( 28 - 3\,z \right) \,\left( 3 - 2\,z \right) \,z\,\log z}
    {9}} +\nn\\
& & {\frac{8\,\left( 1 - z \right) \,\left( 5 + 3\,z \right) \,
      \left( \log (1 - z)\,\log z + 2\,\mathrm{Li_2}(z) \right) }{3}
    }\,,\nn\\
B^u_{4,22}&=&{\frac{-8\,\left( 1 - z \right) \,\left( 11 - 10\,z - 13\,{z^2} \right) }
    {27}} + {\frac{32\,{z^2}\,\log z}{9}}\,,
\eea

\bea
B^{d^{\, \prime}}_{1,11}&=&{\frac{2\,\left( 1 - z \right) \,\left( 2 - z \right) \,
      \left( 5 + 3\,z \right) }{3}} + 
  {\frac{4\,{{\left( 1 - z \right) }^2}\,\left( 4 + 5\,z \right) \,
      \log (1 - z)}{3}} +\nn\\
& & {\frac{4\,z\,\left( 5 + 4\,z - 5\,{z^2} \right) \,
      \log z}{3}} +\nn\\
& & {\frac{2\,{{\left( 1 - z \right) }^2}\,
      \left( 2 + z \right) \,\left( 6\,\log x_0 - 
        4\,\left( \log (1 - z)\,\log z + 2\,\mathrm{Li_2}(z) \right)
            \right) }{3}}\,,\nn\\
B^{d^{\, \prime}}_{2,11}&=&{\frac{-4\,\left( 1 - z \right) \,\left( 5 - 7\,z + 6\,{z^2} \right) }{3}} - 
  {\frac{8\,{{\left( 1 - z \right) }^2}\,
      \left( 2 + 10\,z - 3\,{z^2} \right) \,\log (1 - z)}{3}} + \nn\\
& &  {\frac{8\,z\,\left( 2 - 17\,z + 16\,{z^2} - 3\,{z^3} \right) \,\log z}
    {3}} -\nn\\
& & {\frac{4\,{{\left( 1 - z \right) }^2}\,\left( 1 + 2\,z \right) \,
      \left( 6\,\log x_0 - 
        4\,\left( \log (1 - z)\,\log z + 2\,\mathrm{Li_2}(z) \right)
            \right) }{3}}\,,\nn\\
B^{d^{\, \prime}}_{3,11}&=&{\frac{-\left( \left( 1 - z \right) \,
        \left( 205 + 7\,z - 110\,{z^2} \right)  \right) }{18}} -\nn\\ 
& &  {\frac{3\,{{\left( 1 - z \right) }^2}\,\left( 2 + z \right) \,
      \left(2\, \log x_0 - 2\,\log (1 - z) \right) }{2}} - 
  {\frac{{z^2}\,\left( 6 + 11\,z \right) \,\log z}{3}}\,,\nn\\
B^{d^{\, \prime}}_{4,11}&=&{\frac{\left( 1 - z \right) \,\left( 95 + 104\,z - 211\,{z^2} \right) }{9}} +\nn\\ 
& &  {{\left( 1 - z \right) }^2}\,\left( 1 + 2\,z \right) \,
   \left( 6\,\log x_0 - 6\,\log (1 - z) \right)  - 
  {\frac{4\,\left( 12 - 11\,z \right) \,{z^2}\,\log z}{3}}\,,
\eea

\bea
B^{d^{\, \prime}}_{1,12}&=&{\frac{8\,{{\pi }^2}\,\left( 1 - z \right) \,z\,\left( 2 + z \right) }{27}} + 
  {\frac{2\,\left( 1 - z \right) \,\left( 42 - 15\,z - 25\,{z^2} \right) }
    {9}} +\nn\\
& & {\frac{2\,{{\left( 1 - z \right) }^2}\,\left( 2 + z \right) \,
      \left(2\, \log x_0 + 4\,\log x_1 \right) }{3}} +\nn\\ 
& &  {\frac{4\,{{\left( 1 - z \right) }^2}\,\left( 2 + z + 3\,{z^2} \right) \,
      \log (1 - z)}{9\,z}} + {\frac{4\,z\,
      \left( 1 + 6\,z - 6\,{z^2} \right) \,\log z}{9}} - \nn\\
& &  {\frac{8\,\left( 1 - z \right) \,\left( 2 - z \right) \,
      \left( 2 + z \right) \,\left( \log (1 - z)\,\log z + 
        2\,\mathrm{Li_2}(z) \right) }{9}}\,,\nn\\
B^{d^{\, \prime}}_{2,12}&=&{\frac{-16\,{{\pi }^2}\,\left( 1 - z \right) \,z\,\left( 1 + 2\,z \right) }
    {27}} - {\frac{4\,\left( 1 - z \right) \,
      \left( 21 + 27\,z - 38\,{z^2} \right) }{9}} -\nn\\ 
& &  {\frac{4\,{{\left( 1 - z \right) }^2}\,\left( 1 + 2\,z \right) \,
      \left(2\, \log x_0 + 4\,\log x_1 \right) }{3}} - \nn\\
& &  {\frac{8\,{{\left( 1 - z \right) }^2}\,
      \left( 1 + 2\,z + 6\,{z^2} - 3\,{z^3} \right) \,\log (1 - z)}{9\,z}} +\nn\\
& &  {\frac{8\,z\,\left( 4 - 24\,z + 18\,{z^2} - 3\,{z^3} \right) \,\log z}
    {9}} +\nn\\
& & {\frac{16\,\left( 1 - z \right) \,\left( 2 - z \right) \,
      \left( 1 + 2\,z \right) \,
      \left( \log (1 - z)\,\log z + 2\,\mathrm{Li_2}(z) \right) }{9}
    }\,,\nn\\
B^{d^{\, \prime}}_{3,12}&=&{\frac{2\,{{\pi }^2}\,\left( 1 - z \right) \,z\,\left( 2 + z \right) }{9}} + 
  {\frac{\left( 1 - z \right) \,\left( 83 - 151\,z - 88\,{z^2} \right) }
    {54}} -\nn\\
& & {\frac{{{\left( 1 - z \right) }^2}\,\left( 2 + z \right) \,
      \left(2\, \log x_0 - 4\,\log x_1 \right) }{2}} + \nn\\
& &  {\frac{{{\left( 1 - z \right) }^2}\,\left( 1 + z \right) \,
      \left( 2 + z \right) \,\log (1 - z)}{3\,z}} - 
  {\frac{2\,z\,\left( 6 + 7\,{z^2} \right) \,\log z}{9}} - \nn\\
& &  {\frac{2\,\left( 1 - z \right) \,\left( 2 + z \right) \,
      \left( \log (1 - z)\,\log z + 2\,\mathrm{Li_2}(z) \right) }{3}
    }\,,\nn\\
B^{d^{\, \prime}}_{4,12}&=&{\frac{-4\,{{\pi }^2}\,\left( 1 - z \right) \,z\,\left( 1 + 2\,z \right) }
    {9}} - {\frac{\left( 1 - z \right) \,
      \left( 49 + 202\,z - 185\,{z^2} \right) }{27}} + \nn\\
& &  {{\left( 1 - z \right) }^2}\,\left( 1 + 2\,z \right) \,
   \left(2\, \log x_0 - 4\,\log x_1 \right)  -\nn\\ 
& &  {\frac{2\,{{\left( 1 - z \right) }^2}\,\left( 1 + z \right) \,
      \left( 1 + 2\,z \right) \,\log (1 - z)}{3\,z}} + 
  {\frac{2\,z\,\left( 6 - 45\,z + 28\,{z^2} \right) \,\log z}{9}} +\nn\\ 
& &  {\frac{4\,\left( 1 - z \right) \,\left( 1 + 2\,z \right) \,
      \left( \log (1 - z)\,\log z + 2\,\mathrm{Li_2}(z) \right) }{3}
    }\,,
\eea

\bea
B^{d^{\, \prime}}_{1,22}&=&{\frac{16\,{{\pi }^2}\,\left( 1 - z \right) \,z\,\left( 2 + z \right) }
    {81}} + {\frac{2\,\left( 1 - z \right) \,
      \left( 461 - 286\,z - 313\,{z^2} \right) }{243}} +\nn\\ 
& &  {\frac{16\,{{\left( 1 - z \right) }^2}\,\left( 2 + z \right) \,
      \log x_1}{9}} + 
  {\frac{16\,{{\left( 1 - z \right) }^2}\,\left( 1 + z + {z^2} \right) \,
      \log (1 - z)}{27\,z}} -\nn\\
& & {\frac{4\,z\,
      \left( 9 - 18\,z + 32\,{z^2} \right) \,\log z}{81}} -\nn\\ 
& &  {\frac{8\,\left( 1 - z \right) \,\left( 3 - z \right) \,
      \left( 2 + z \right) \,\left( \log (1 - z)\,\log z + 
        2\,\mathrm{Li_2}(z) \right) }{27}}\,,\nn\\
B^{d^{\, \prime}}_{2,22}&=&{\frac{-32\,{{\pi }^2}\,\left( 1 - z \right) \,z\,\left( 1 + 2\,z \right) }
    {81}} - {\frac{4\,\left( 1 - z \right) \,
      \left( 238 + 445\,z - 527\,{z^2} \right) }{243}} -\nn\\ 
& &  {\frac{32\,{{\left( 1 - z \right) }^2}\,\left( 1 + 2\,z \right) \,
      \log x_1}{9}} - \nn\\
& &  {\frac{8\,{{\left( 1 - z \right) }^2}\,
      \left( 2 + 5\,z + 8\,{z^2} - 3\,{z^3} \right) \,\log (1 - z)}{27\,z}} +\nn\\ 
& &  {\frac{8\,z\,\left( 18 - 117\,z + 82\,{z^2} - 9\,{z^3} \right) \,\log z}
    {81}} +\nn\\
& & {\frac{16\,\left( 1 - z \right) \,\left( 3 - z \right) \,
      \left( 1 + 2\,z \right) \,
      \left( \log (1 - z)\,\log z + 2\,\mathrm{Li_2}(z) \right) }{
      27}}\,,\nn\\
B^{d^{\, \prime}}_{3,22}&=&{\frac{2\,{{\pi }^2}\,\left( 1 - z \right) \,\left( 2 + z \right) \,
      \left( 9 + 7\,z \right) }{27}} - 
  {\frac{\left( 1 - z \right) \,\left( 937 - 449\,z - 746\,{z^2} \right) }
    {81}} -\nn\\
& & {\frac{4\,{{\left( 1 - z \right) }^2}\,\left( 2 + z \right) \,
      \log x_1}{3}} - 
  {\frac{4\,{{\left( 1 - z \right) }^2}\,
      \left( 1 - 17\,z - 8\,{z^2} \right) \,\log (1 - z)}{9\,z}} +\nn\\ 
& &  {\frac{2\,z\,\left( 72 - 9\,z - 20\,{z^2} \right) \,\log z}{27}} - \nn\\
& &  {\frac{4\,\left( 1 - z \right) \,\left( 2 + z \right) \,
      \left( 3 + 5\,z \right) \,
      \left( \log (1 - z)\,\log z + 2\,\mathrm{Li_2}(z) \right) }{9}
    }\,,\nn\\
B^{d^{\, \prime}}_{4,22}&=&{\frac{-4\,{{\pi }^2}\,\left( 1 - z \right) \,\left( 1 + 2\,z \right) \,
      \left( 9 + 7\,z \right) }{27}} + 
  {\frac{\left( 1 - z \right) \,\left( 715 + 1003\,z - 2804\,{z^2} \right) }
    {81}} +\nn\\
& & {\frac{8\,{{\left( 1 - z \right) }^2}\,\left( 1 + 2\,z \right) \,
      \log x_1}{3}} +\nn\\ 
& &  {\frac{2\,{{\left( 1 - z \right) }^2}\,
      \left( 2 - 31\,z - 64\,{z^2} - 3\,{z^3} \right) \,\log (1 - z)}{9\,z}} +\nn\\
& &   {\frac{2\,z\,\left( 36 - 288\,z + 62\,{z^2} + 9\,{z^3} \right) \,\log z}
    {27}} +\nn\\
& & {\frac{8\,\left( 1 - z \right) \,\left( 1 + 2\,z \right) \,
      \left( 3 + 5\,z \right) \,
      \left( \log (1 - z)\,\log z + 2\,\mathrm{Li_2}(z) \right) }{9}
    }\,,
\eea

\bea
B^{s^{\, \prime}}_{1,11}&=&{\frac{4\,{\sqrt{1 - 4\,z}}\,\left( 1 - z \right) \,\left( 5 + 6\,z \right) }
    {3}} + {\frac{8\,\left( 6 - 13\,z - 2\,{z^2} + 6\,{z^3} \right) \,
      \log \sigma}{3}} +\nn\\
& & {\frac{4\,{\sqrt{1 - 4\,z}}\,\left( 1 - z \right) \,
      \left( 6\,\log x_0 + 8\,\log (1 - 4\,z) - 12\,\log z
         \right) }{3}} -\nn\\
& & \frac{16}{3}\,\left( 1 - 2\,z \right) \,
      \left( 1 - z \right) \,\left( 3\,\log^2 \sigma + 
        2\,\log \sigma\,\log (1 - 4\,z) - 3\,\log \sigma\,\log z + 
        4\,\mathrm{Li_2}(\sigma) + \right. \nn\\
& &  \left. 2\,\mathrm{Li_2}(\sigma^2) \right)\,,\nn\\
B^{s^{\, \prime}}_{2,11}&=&{- \frac{4\,{\sqrt{1 - 4\,z}}\,\left( 1 - 4\,z \right) \,
      \left( 5 + 6\,z \right) }{3}} - 
  {\frac{16\,\left( 3 - 2\,z - 7\,{z^2} + 12\,{z^3} \right) \,\log \sigma}{3}} + \nn\\
& &  16\,\left( 1 - 2\,z \right) \,\left( 1 + 2\,z \right) \,\log^2 \sigma + 
  {\frac{32\,\left( 1 - 2\,z \right) \,\left( 1 + 2\,z \right) \,\log \sigma\,
      \log (1 - 4\,z)}{3}} -\nn\\
& & {\frac{4\,{\sqrt{1 - 4\,z}}\,
      \left( 1 + 2\,z \right) \,
      \left( 6\,\log x_0 + 8\,\log (1 - 4\,z) - 12\,\log z
         \right) }{3}} -\nn\\
& & 16\,\left( 1 - 2\,z \right) \,
   \left( 1 + 2\,z \right) \,\log \sigma\,\log z + 
  {\frac{64\,\left( 1 - 2\,z \right) \,\left( 1 + 2\,z \right) \,
      \mathrm{Li_2}(\sigma)}{3}} +\nn\\ 
& &  {\frac{32\,\left( 1 - 2\,z \right) \,\left( 1 + 2\,z \right) \,
      \mathrm{Li_2}({\sigma^2})}{3}}\,,\nn\\
B^{s^{\, \prime}}_{3,11}&=&{\frac{-\left( {\sqrt{1 - 4\,z}}\,\left( 205 + 14\,z + 24\,{z^2} \right) 
         \right) }{18}} + {\frac{2\,
      \left( 9 - 27\,z - 6\,{z^2} + 4\,{z^3} \right) \,\log \sigma}{3}} - \nn\\
& &  3\,{\sqrt{1 - 4\,z}}\,\left( 1 - z \right) \,
   \left(2\, \log x_0 - 2\,\log (1 - 4\,z) + 2\,\log z \right)\,,\nn\\ 
B^{s^{\, \prime}}_{4,11}&=&{\frac{{\sqrt{1 - 4\,z}}\,\left( 95 + 208\,z + 48\,{z^2} \right) }{9}} - 
  {\frac{2\,\left( 9 - 6\,{z^2} + 16\,{z^3} \right) \,\log \sigma}{3}} +\nn\\& &  3\,{\sqrt{1 - 4\,z}}\,\left( 1 + 2\,z \right) \,
   \left(2\, \log x_0 - 2\,\log (1 - 4\,z) + 2\,\log z \right)\,, 
\eea

\bea
B^{s^{\, \prime}}_{1,12}&=&{\frac{{\sqrt{1 - 4\,z}}\,\left( 26 - 17\,z - 6\,{z^2} \right) }{3}} - 
  {\frac{\left( 1 - 60\,z + 146\,{z^2} - 36\,{z^4} \right) \,\log \sigma}
    {9\,z}} +\nn\\
& & {\frac{4\,{\sqrt{1 - 4\,z}}\,\left( 1 - z \right) \,
      \left( 6\,\log x_0 + 12\,\log x_1 + 
        8\,\log (1 - 4\,z) \right) }{9}} + \nn\\
& &  {\frac{{\sqrt{1 - 4\,z}}\,\left( 1 - 58\,z + 72\,{z^2} \right) \,\log z}
    {9\,z}} -\nn\\
& & \frac{16}{9}\,\left( 1 - 2\,z \right) \,\left( 1 - z \right) \,
      \left( 3\,\log^2 \sigma + 2\,\log \sigma\,\log (1 - 4\,z) - 
        3\,\log \sigma\,\log z + 4\,\mathrm{Li_2}(\sigma) + \right. \nn \\
& &   \left. 2\,\mathrm{Li_2}({\sigma^2}) \right) \,,\nn\\
B^{s^{\, \prime}}_{2,12}&=&{\frac{-4\,{\sqrt{1 - 4\,z}}\,\left( 7 - 3\,z \right) \,
      \left( 1 + 2\,z \right) }{3}} + 
  {\frac{4\,\left( 1 - 15\,z - 4\,{z^2} + 48\,{z^3} - 36\,{z^4} \right) \,
      \log \sigma}{9\,z}} -\nn\\
& & {\frac{4\,{\sqrt{1 - 4\,z}}\,
      \left( 1 + 2\,z \right) \,
      \left( 6\,\log x_0 + 12\,\log x_1 + 
        8\,\log (1 - 4\,z) \right) }{9}} - \nn\\
& &  {\frac{4\,{\sqrt{1 - 4\,z}}\,\left( 1 - 13\,z - 36\,{z^2} \right) \,
      \log z}{9\,z}} +\nn\\
& & \frac{16 }{9}\,\left( 1 - 2\,z \right) \,
      \left( 1 + 2\,z \right) \,
      \left( 3\,\log^2 \sigma + 2\,\log \sigma\,\log (1 - 4\,z) - 
        3\,\log \sigma\,\log z + 4\,\mathrm{Li_2}(\sigma) + \right. \nn \\
& &  \left. 2\,\mathrm{Li_2}({\sigma^2}) \right)\,,\nn\\
B^{s^{\, \prime}}_{3,12}&=&{\frac{{\sqrt{1 - 4\,z}}\,\left( 112 - 523\,z + 6\,{z^2} \right) }{108}} - 
  {\frac{\left( 3 - 108\,z + 342\,{z^2} + 4\,{z^4} \right) \,\log \sigma}
    {36\,z}} -\nn\\
& & {\sqrt{1 - 4\,z}}\,\left( 1 - z \right) \,
   \left(2\, \log x_0 - 4\,\log x_1 - 
     2\,\log (1 - 4\,z) \right)  +\nn\\ 
& &  {\frac{{\sqrt{1 - 4\,z}}\,\left( 1 - 34\,z + 48\,{z^2} \right) \,\log z}
    {12\,z}}\,,\nn\\
B^{s^{\, \prime}}_{4,12}&=&{\frac{-\left( {\sqrt{1 - 4\,z}}\,\left( 49 + 215\,z + 6\,{z^2} \right) 
         \right) }{27}} + {\frac{\left( 3 - 27\,z - 36\,{z^2} + 72\,{z^3} + 
        4\,{z^4} \right) \,\log \sigma}{9\,z}} +\nn\\ 
& &  {\sqrt{1 - 4\,z}}\,\left( 1 + 2\,z \right) \,
   \left(2\, \log x_0 - 4\,\log x_1 - 
     2\,\log (1 - 4\,z) \right)  -\nn\\ 
& &  {\frac{{\sqrt{1 - 4\,z}}\,\left( 1 - 7\,z - 24\,{z^2} \right) \,\log z}
    {3\,z}}\,,
\eea

\bea
B^{s^{\,\prime}}_{1,22}&=&{\frac{2\,{\sqrt{1 - 4\,z}}\,\left( 407 - 491\,z - 78\,{z^2} \right) }
    {243}} - {\frac{2\,\left( 3 - 144\,z + 390\,{z^2} - 52\,{z^4} \right) \,
      \log \sigma}{81\,z}} +\nn\\
& & {\frac{8\,{\sqrt{1 - 4\,z}}\,\left( 1 - z \right) \,
      \left( 12\,\log x_1 + 7\,\log (1 - 4\,z) \right) }{27}} + 
  {\frac{2\,{\sqrt{1 - 4\,z}}\,\left( 1 - 46\,z + 60\,{z^2} \right) \,
      \log z}{27\,z}} -\nn\\
& & \frac{16}{27}\,\left( 1 - 2\,z \right) \,
      \left( 1 - z \right) \,\left( 3\,\log^2 \sigma + 
        2\,\log \sigma\,\log (1 - 4\,z) - 3\,\log \sigma\,\log z + 
        4\,\mathrm{Li_2}(\sigma) + \right. \nn \\
& & \left. 2\,\mathrm{Li_2}({\sigma^2}) \right) \,,\nn\\
B^{s^{\,\prime}}_{2,22}&=&{\frac{-8\,{\sqrt{1 - 4\,z}}\,\left( 119 + 256\,z - 78\,{z^2} \right) }
    {243}} + {\frac{8\,\left( 3 - 36\,z - 24\,{z^2} + 108\,{z^3} - 
        52\,{z^4} \right) \,\log \sigma}{81\,z}} -\nn\\ 
& &  {\frac{8\,{\sqrt{1 - 4\,z}}\,\left( 1 + 2\,z \right) \,
      \left( 12\,\log x_1 + 7\,\log (1 - 4\,z) \right) }{27}} -
 {\frac{8\,{\sqrt{1 - 4\,z}}\,\left( 1 - 10\,z - 30\,{z^2} \right) \,
      \log z}{27\,z}} +\nn\\
& & \frac{16}{27}\,\left( 1 - 2\,z \right) \,
      \left( 1 + 2\,z \right) \,
      \left( 3\,\log^2 \sigma + 2\,\log \sigma\,\log (1 - 4\,z) - 
        3\,\log \sigma\,\log z + 4\,\mathrm{Li_2}(\sigma) + \right. \nn \\
& &  \left.  2\,\mathrm{Li_2}({\sigma^2}) \right)\,,\nn\\
B^{s^{\,\prime}}_{3,22}&=&{\frac{4\,{{\pi }^2}\,\left( 1 - z \right) }{3}} - 
  {\frac{{\sqrt{1 - 4\,z}}\,\left( 1129 - 2143\,z + 186\,{z^2} \right) }
    {81}} +\nn\\
& & {\frac{2\,\left( 21 + 78\,z - 267\,{z^2} + 90\,{z^3} + 
        62\,{z^4} \right) \,\log \sigma}{27\,z}} - \nn\\
& &  {\frac{8\,\left( 2 - 5\,z \right) \,{\sqrt{1 - 4\,z}}\,
      \log x_1}{9}} + 
  {\frac{112\,{\sqrt{1 - 4\,z}}\,\left( 1 - z \right) \,\log (1 - 4\,z)}
    {9}} -\nn\\
& & {\frac{4\,\left( 1 - z \right) \,\left( 7 + 4\,z \right) \,
      \log \sigma\,\log (1 - 4\,z)}{9}} - 
  {\frac{2\,{\sqrt{1 - 4\,z}}\,\left( 7 + 40\,z - 71\,{z^2} \right) \,
      \log z}{9\,z}} -\nn\\
& & {\frac{4\,\left( 1 - z \right) \,
      \left( 5 + 2\,z \right) \,
      \left( \log^2 \sigma - \log \sigma\,\log z \right) }{3}} +\nn\\& & {\frac{16\,\left( 1 - 2\,z \right) \,\left( 1 - z \right) \,
      \mathrm{Li_2}(\sigma)}{9}} - 
  {\frac{16\,\left( 1 - z \right) \,\left( 4 + z \right) \,
      \mathrm{Li_2}({\sigma^2})}{9}}\,,\nn\\
B^{s^{\,\prime}}_{4,22}&=&{\frac{-4\,{{\pi }^2}\,\left( 1 + 2\,z \right) }{3}} + 
  {\frac{{\sqrt{1 - 4\,z}}\,\left( 691 + 1922\,z + 1608\,{z^2} \right) }
    {81}} -\nn\\
& & {\frac{2\,\left( 3 + 159\,z + 138\,{z^2} - 774\,{z^3} + 
        536\,{z^4} \right) \,\log \sigma}{27\,z}} + \nn\\
& &  {\frac{8\,{\sqrt{1 - 4\,z}}\,\left( 1 + 2\,z \right) \,
      \left(2\,\log x_1 - 14\,\log (1 - 4\,z) \right) }{9}} + \nn\\
& &  {\frac{4\,\left( 1 + 2\,z \right) \,\left( 7 + 4\,z \right) \,\log \sigma\,
      \log (1 - 4\,z)}{9}} + {\frac{2\,{\sqrt{1 - 4\,z}}\,
      \left( 1 + 55\,z + 154\,{z^2} \right) \,\log z}{9\,z}} +\nn\\ 
& &  {\frac{4\,\left( 1 + 2\,z \right) \,\left( 5 + 2\,z \right) \,
      \left( \log^2 \sigma - \log \sigma\,\log z \right) }{3}} - \nn\\
& &  {\frac{16\,\left( 1 - 2\,z \right) \,\left( 1 + 2\,z \right) \,
      \mathrm{Li_2}(\sigma)}{9}} + 
  {\frac{16\,\left( 4 + z \right) \,\left( 1 + 2\,z \right) \,
      \mathrm{Li_2}({\sigma^2})}{9}}\,,
\eea
\end{footnotesize}
\renewcommand{\arraystretch}{1.0} 
\noindent where $\sigma$ is the ratio
\be
\sigma = \frac{1-\sqrt{1-4z}}{1+\sqrt{1-4z}} \, ,
\ee
and we have defined $x_0=\mu_0/m_b$ and $x_1=\mu_1/m_b$.

Finally we present the results for the coefficients $P_{k,ij}$ of the penguin
and chromomagnetic operators defined in eq.~(\ref{eq:c6sdb0}). The 
coefficients $P_{k,28}$ have been computed by using the convention in which 
the chromomagnetic coefficient $C_{8G}$ has a positive sign. We obtain the 
expressions:
\begin{footnotesize}
\be
\begin{array}{lll}
  P_{1,13}=-{\sqrt{1 - 4\,z}}\,\left( 1 - z \right) \,, \qquad
& P_{1,23}={\dfrac{-{\sqrt{1 - 4\,z}}\,\left( 1 - z \right)}{3}}\,,\qquad
& P_{1,14}={\dfrac{-{\sqrt{1 - 4\,z}}\,\left( 1 - z \right)}{3}}\,,\\
  P_{2,13}={\sqrt{1 - 4\,z}}\,\left( 1 + 2\,z \right)\,, \qquad
& P_{2,23}={\dfrac{{\sqrt{1 - 4\,z}}\,\left( 1 + 2\,z \right) }{3}}\,, \qquad
& P_{2,14}={\dfrac{{\sqrt{1 - 4\,z}}\,\left( 1 + 2\,z \right) }{3}}\,, \\
  P_{3,13}=0\,, \qquad
& P_{3,23}=0\,, \qquad
& P_{3,14}=0\,,\\
  P_{4,13}=0\,, \qquad
& P_{4,23}=0\,, \qquad
& P_{4,14}=0\,. 
\end{array}
\ee

\be
\begin{array}{lll}
  P_{1,24}={\dfrac{-{\sqrt{1 - 4\,z}}\,\left( 1 - z \right)}{9}}\,, \qquad
& P_{1,15}=-3\,{z\,\sqrt{1 - 4\,z}}\,,\qquad
& P_{1,25}=-{z\,\sqrt{1 - 4\,z}}\,,\\
  P_{2,24}={\dfrac{{\sqrt{1 - 4\,z}}\,\left( 1 + 2\,z \right) }{9}}\,, \qquad
& P_{2,15}=0\,, \qquad
& P_{2,25}=0\,, \\
  P_{3,24}={\dfrac{-2\,{\sqrt{1 - 4\,z}}\,\left( 1 - z \right) }{3}}\,, \qquad
& P_{3,15}=0\,, \qquad
& P_{3,25}=0\,,\\
  P_{4,24}={\dfrac{2\,{\sqrt{1 - 4\,z}}\,\left( 1 + 2\,z \right) }{3}}\,, \qquad
& P_{4,15}=0\,, \qquad
& P_{4,25}=0\,. 
\end{array}
\ee

\be
\begin{array}{lll}
  P_{1,16}=-{z\,\sqrt{1 - 4\,z}}\,, \qquad
& P_{1,26}={\dfrac{-{z\,\sqrt{1 - 4\,z}}}{3}}\,,\qquad
& P_{1,28}=0\,,\\
  P_{2,16}=0\,, \qquad
& P_{2,26}=0\,, \qquad
& P_{2,28}=0\,, \\
  P_{3,16}=0\,, \qquad
& P_{3,26}=-2\,{z\,\sqrt{1 - 4\,z}}\,, \qquad
& P_{3,28}={\dfrac{-2\,{\sqrt{1 - 4\,z}}\,\left( 1 + 2\,z \right) }{3}}\,,\\
  P_{4,16}=0\,, \qquad
& P_{4,26}=0\,, \qquad
& P_{4,28}={\dfrac{2\,{\sqrt{1 - 4\,z}}\,\left( 1 + 2\,z \right) }{3}}\,. 
\end{array}
\ee
\end{footnotesize}
Note that, in the limit of vanishing charm quark mass $(z=0)$, the contribution
of the penguin operators $Q_5$ and $Q_6$ vanish for chirality.



\begin{thebibliography}{99}

\bibitem{Khoze:1987fa}
V.A.~Khoze, M.A.~Shifman, N.G.~Uraltsev and M.B.~Voloshin,
Sov.\ J.\ Nucl.\ Phys.\  {\bf 46} (1987) 112
[Yad.\ Fiz.\  {\bf 46} (1987) 181].

\bibitem{ope}
J.~Chay, H.~Georgi and B.~Grinstein,
Phys.\ Lett.\ B {\bf 247} (1990) 399.

\bibitem{primisp}
I.~I.~Bigi, N.~G.~Uraltsev and A.~I.~Vainshtein,
Phys.\ Lett.\ B {\bf 293} (1992) 430
[Erratum-ibid.\ B {\bf 297} (1992) 477]
[hep-ph/9207214].

\bibitem{NS}
M.~Neubert and C.T.~Sachrajda,
Nucl.\ Phys.\ B {\bf 483}, 339 (1997)
[hep-ph/9603202].

\bibitem{noantri}
M.~Ciuchini, E.~Franco, V.~Lubicz and F.~Mescia,
hep-ph/0110375.

\bibitem{bucha}
M.~Beneke, G.~Buchalla, C.~Greub, A.~Lenz and U.~Nierste,
hep-ph/0202106.

\bibitem{DiPierro98}
M.~Di Pierro and C.T.~Sachrajda  [UKQCD Collaboration],
Nucl.\ Phys.\ B {\bf 534} (1998) 373
[hep-lat/9805028].

\bibitem{DiPierro:1999tb}
M.~Di Pierro, C.~T.~Sachrajda and C.~Michael  [UKQCD collaboration],
Phys.\ Lett.\ B {\bf 468} (1999) 143
[hep-lat/9906031].

\bibitem{DiPierroproc}
M.~Di Pierro and C.T.~Sachrajda  [UKQCD collaboration],
Nucl.\ Phys.\ Proc.\ Suppl.\  {\bf 73} (1999) 384
[hep-lat/9809083].

\bibitem{APE} D.~Becirevic, private communication; updated results with respect
to hep-ph/0110124

\bibitem{ckmwork}
E.~Barberio, presented at the Worksop on the CKM Unitarity Triangle,
CERN, Geneve, February 13-16, 2002, http://ckm-workshop.web.cern.ch/ 

\bibitem{Bloch:1937pw}
F.~Bloch and A.~Nordsieck,
Phys.\ Rev.\ {\bf 52} (1937) 54.

\bibitem{Doria:ak}
R.~Doria, J.~Frenkel and J.~C.~Taylor,
Nucl.\ Phys.\ B {\bf 168}, 93 (1980).

\bibitem{Di'Lieto:1980dt}
C.~Di'Lieto, S.~Gendron, I.~G.~Halliday and C.~T.~Sachrajda,
Nucl.\ Phys.\ B {\bf 183}, 223 (1981).

\bibitem{nlodb1a}
A.J.~Buras, M.~Jamin, M.E.~Lautenbacher and P.H.~Weisz,
Nucl.\ Phys.\ B {\bf 400} (1993) 37
[hep-ph/9211304];

\bibitem{nlodb1b}
A.J.~Buras, M.~Jamin and M.E.~Lautenbacher,
Nucl.\ Phys.\ B {\bf 400} (1993) 75
[hep-ph/9211321];

\bibitem{nlodb1c}
M.~Ciuchini, E.~Franco, G.~Martinelli and L.~Reina,
Nucl.\ Phys.\ B {\bf 415} (1994) 403
[hep-ph/9304257].

\bibitem{Bigi:1992su}
I.I.~Bigi, N.G.~Uraltsev and A.I.~Vainshtein,
Phys.\ Lett.\ B {\bf 293}, 430 (1992); Erratum {\bf 297}, 477 (1993)
[hep-ph/9207214].

\bibitem{gamma1}
Q.~Ho-kim and X.-y.~Pham,
Phys.\ Lett.\ B {\bf 122} (1983) 297;

\bibitem{gamma2}
Y.~Nir,
Phys.\ Lett.\ B {\bf 221} (1989) 184;

\bibitem{gamma3}
E.~Bagan, P.~Ball, V.M.~Braun and P.~Gosdzinsky,
Nucl.\ Phys.\ B {\bf 432} (1994) 3
[hep-ph/9408306];

\bibitem{gamma4}
E.~Bagan, P.~Ball, V.M.~Braun and P.~Gosdzinsky,
Phys.\ Lett.\ B {\bf 342} (1995) 362; Erratum {\bf 374} (1996) 363
[hep-ph/9409440];

\bibitem{gamma5}
E.~Bagan, P.~Ball, B.~Fiol and P.~Gosdzinsky,
Phys.\ Lett.\ B {\bf 351} (1995) 546
[hep-ph/9502338];

\bibitem{gamma6}
A.F.~Falk, Z.~Ligeti, M.~Neubert and Y.~Nir,
Phys.\ Lett.\ B {\bf 326} (1994) 145
[hep-ph/9401226].

\bibitem{Ura}
N.~G.~Uraltsev,
Phys.\ Lett.\ B {\bf 376} (1996) 303
[hep-ph/9602324].

\bibitem{charm} 
M.~Beneke and G.~Buchalla,
Phys.\ Rev.\ D {\bf 53}, 4991 (1996)
[hep-ph/9601249].

\bibitem{integrals1}
K.G.~Chetyrkin and F.V.~Tkachov,
Nucl.\ Phys.\ B {\bf 192} (1981) 159;

\bibitem{integrals2}
O.V.~Tarasov,
Nucl.\ Phys.\ B {\bf 502} (1997) 455
[hep-ph/9703319];

\bibitem{integrals3}
O.~V.~Tarasov,
Phys.\ Rev.\ D {\bf 54} (1996) 6479
[hep-th/9606018].

\bibitem{tarcer}
R.~Mertig and R.~Scharf,
Comput.\ Phys.\ Commun.\  {\bf 111} (1998) 265
[hep-ph/9801383].

\bibitem{Misiak:1999yg}
M.~Misiak and J.~Urban,
Phys.\ Lett.\ B {\bf 451} (1999) 161
[hep-ph/9901278].

\bibitem{reyes}
V.~Gimenez and J.~Reyes,
Nucl.\ Phys.\ B {\bf 545}, 576 (1999)
[hep-lat/9806023].

\bibitem{Beneke:1999sy}
M.~Beneke, G.~Buchalla, C.~Greub, A.~Lenz and U.~Nierste,
Phys.\ Lett.\ B {\bf 459}, 631 (1999)
[hep-ph/9808385].

\bibitem{db2nlo1}
M.~Ciuchini, E.~Franco, V.~Lubicz, G.~Martinelli, I.~Scimemi and L.~Silvestrini,
Nucl.\ Phys.\ B {\bf 523}, 501 (1998)
[hep-ph/9711402];

\bibitem{mupig}
M.~Neubert,
Adv.\ Ser.\ Direct.\ High Energy Phys.\  {\bf 15} (1998) 239
[hep-ph/9702375].

\bibitem{Chernyak:1995cx}
V.~Chernyak,
Nucl.\ Phys.\ B {\bf 457} (1995) 96
[hep-ph/9503208].

\bibitem{pirjol}
D.~Pirjol and N.~Uraltsev,
Phys.\ Rev.\ D {\bf 59}, 034012 (1999)
[hep-ph/9805488].

\bibitem{Colangelo:1996ta}
P.~Colangelo and F.~De Fazio,
Phys.\ Lett.\ B {\bf 387} (1996) 371
[hep-ph/9604425].

\bibitem{Baek:1998vk}
M.~S.~Baek, J.~Lee, C.~Liu and H.~S.~Song,
Phys.\ Rev.\ D {\bf 57}, 4091 (1998) [hep-ph/9709386].

\bibitem{Cheng:1999ia}
H.~Y.~Cheng and K.~C.~Yang,
Phys.\ Rev.\ D {\bf 59} (1999) 014011
[hep-ph/9805222].

\bibitem{Huang:1999xj}
C.~S.~Huang, C.~Liu and S.~L.~Zhu,
Phys.\ Rev.\ D {\bf 61}, 054004 (2000)
[hep-ph/9906300].

\end{thebibliography}
\end{document}